\documentclass[aps,prb, 11pt, superscriptaddress,notitlepage]{revtex4-1}

\usepackage{pdfpages}
\usepackage{graphicx}
\usepackage{amsmath}
\usepackage{subfig}
\usepackage{mathtools}
\usepackage{comment}
\usepackage{amsmath}
\usepackage{amssymb}
\usepackage[below]{placeins}
\usepackage{fancyhdr}
\usepackage{bm}

\newcommand{\beginsupplement}{%
        \setcounter{section}{0}
        \renewcommand{\thesection}{Supplementary Note \arabic{section}}%
        \setcounter{table}{0}
         \renewcommand{\tablename}{}
        \renewcommand{\thetable}{Supplementary Table \arabic{table}}
        \setcounter{figure}{0}
         \renewcommand{\figurename}{}
        \renewcommand{\thefigure}{Supplementary Figure \arabic{figure}}
     }

\definecolor{LightCyan}{RGB}{240,250,0}

\makeatletter
\AtBeginDocument{\let\LS@rot\@undefined}
\makeatother

\begin{document}



# Induced unconventional superconductivity on the surface states of Bi$_2$Te$_3$ topological insulator

## Authors:

S. Charpentier[1†], L. Galletti[1†], G. Kunakova[1,2], R. Arpaia[1], Y. Song[1,3], R. Baghdadi[1], S. M. Wang[1,3], A. Kalaboukhov[1], E. Olsson[4], F. Tafuri[5], D. Golubev[6], J. Linder[7], T. Bauch[1] and F. Lombardi[1*]

## Affiliations:

[1]Department of Microtechnology and Nanoscience, Chalmers University of Technology, SE-41296 Göteborg, Sweden

[2]Institute of Chemical Physics, University of Latvia, 19 Raina Blvd., LV-1586, Riga, Latvia

[3]Shanghai Institute of Microsystem and Information Technology, Chinese Academy of Sciences, 865 Changning Road, Shanghai CN-200050, China

[4]Department of Applied Physics, Chalmers University of Technology, SE-41296 Göteborg, Sweden

[5]Dipartimento di Ingegneria dell'Informazione, Seconda Università di Napoli, IT-81031 Aversa (CE), Italy and CNR-SPIN

[6]Department of Applied Physics, Aalto University School of Science, P.O. Box 13500, FI-00076 Aalto, Finland

[7]Department of Physics, Norwegian University of Science and Technology, N-7491 Trondheim, Norway

*Correspondence to: floriana.lombardi@chalmers.se

† These authors equally contributed to the work

## Abstract:

Topological superconductivity is central to a variety of novel phenomena involving the interplay between topologically ordered phases and broken-symmetry states. The key ingredient is an unconventional order parameter (OP), with an orbital component containing a chiral $p_x$+i$p_y$ wave term. Here we present phase-sensitive measurements, based on the quantum interference in nanoscale Josephson junctions, realized by using Bi$_2$Te$_3$ topological insulator. We demonstrate that the induced superconductivity is unconventional and consistent with a sign-changing order parameter, such as a chiral $p_x$+i$p_y$ component. The magnetic field pattern of the junctions shows a dip at zero externally applied magnetic field, which is an incontrovertible signature of the simultaneous existence of 0 and π coupling within the junction, inherent to a non trivial OP phase. The nano-textured morphology of the Bi$_2$Te$_3$ flakes, and the dramatic role played by thermal strain are the surprising key factors for the display of an unconventional induced OP.

# Introduction:

The notion of superconducting materials with fully gapped order parameters, but that still support topologically protected surface states[1-5], including Majorana bound states[2,3], are currently subject to much attention. Such systems intertwine two key paradigms in condensed matter physics, namely spontaneous symmetry breaking and topology, which has been predicted to cause unusual quantum phenomena with even direct analogies to concepts in high-energy physics, such as axion couplings[6]. This novel physics has been recently predicted to occur also in systems where the superconducting phenomenon involves topological Dirac electrons and it can be induced by the proximity with a conventional superconductor.

Heterostructures involving a conventional superconductor (S) and an exotic conductor, represented by the surface of a 3D topological insulator[7-11] (TI) and the edge states of two-dimensional quantum wells[12,13], are ideal systems to emulate topological superconductivity. In the pioneering work by Fu and Kane[1] the superconductivity induced in the surface states of a 3D TI is described, in a basis where the operators acquire a phase factor, by an OP which resembles a spinless chiral $p_x+ip_y$ ($p$) but does not break time reversal symmetry. In a subsequent work, Tkatchov at al.[14] have derived, this time in a canonical operator basis, that the induced superconductivity has an orbital term of the type $p+s$-wave with the $p$ part, consisting of conventional chiral $p_x+ip_y$ and $p_x-ip_y$, that sums up with the $s$-wave term through Pauli matrices to form the total superconducting order parameter.

It is clear that the first step that must be taken to unveil the rich variety of the predicted phenomena related to topological superconductivity is to clearly demonstrate the

unconventional nature of the OP, which identifies in a chiral $p_x$+i$p_y$ ($p$-wave) orbital term. In compounds like the $Sr_2RuO_4$ layered perovskite, for example, considered the archetype of a topological superconductor, the existence of a chiral $p_x$+i$p_y$ wave orbital order, with a nontrivial internal phase, is still not generally accepted[15]. The appearance of Majorana fermions and topological quantum computation are fundamentally encoded in the properties of such $p$-wave superconductivity.

Josephson interferometry emerges as a crucial tool to show key features of the superconductivity involving topological Dirac electrons, as for instance supercurrents induced in the helical edge states of HgTe/HgCdTe[12] and InAs/GaSb[13] 2D TI quantum wells. The helical metallic surface states of 3D TI, probed by Josephson interferometry have shown, up to now, a mostly conventional induced superconductivity. Here the Josephson transport has been demonstrated to be ballistic[8,10] and uniquely associated with the helical surface states[16] which rules out a contribution from the disordered bulk[8,10]. However no other signatures have been observed. The reports on a skewed Josephson current phase relation could not be clearly discriminated if due to a conventional high transparency barrier in S-3DTI-S junctions[17] or to the peculiar bound state spectrum of the surface helical metal[11,14] possibly hosting Majorana fermions.

Our experiment is a decisive step along this line. We have realized hybrid devices based on a 3D TI, where the nanostructured morphology of the topological insulator flakes makes the measurement of the Josephson effect sensitive to the interface unconventional order parameter symmetry. By using the same path and notions previously employed to prove the $d$-wave OP in high critical temperature superconductors[18,19] (HTS) we have measured an unconventional magnetic field pattern of the Josephson current, with a dip at zero magnetic

field. We demonstrate that this is the incontrovertible code of a Cooper pair tunneling process probing an OP with a non-trivial internal phase.

# Results:

### Unconventional magnetic field patterns in Al-Bi$_2$Te$_3$-Al Josephson junctions.

Al-3DTI-Al Josephson junctions (Fig. 1a) have been realized using Bi$_2$Te$_3$ flakes from epitaxial thin films. By varying the interface transparency $D$, from a value close to 1 to the tunneling limit ($D \ll 1$) we have been able to show that the surface states of Bi$_2$Te$_3$ host an induced OP with a non trivial phase compatible with a chiral $p$-wave. Figure 1b shows the nanostructured morphology of the flakes. We observe characteristic pyramidal domains (see Methods section) with a prevailing single domain, which points towards thin films without twin boundaries[20].

Figure 1c shows the current voltage characteristic (IVC) of a typical junction at the first cool down, measured at base temperature of 20 mK in a dilution refrigerator. The curve presents 50% hysteresis and a large excess current $I_{exc}$, defined as the extrapolation at zero voltage of the IVC above the Al gap, $\Delta_{Al}$. This value can be higher than the Josephson current for high transparent interfaces[21] as we generally find for our devices. We will assume that the junctions can be assimilated to S'INIS' structures where S' is the effective fully gapped $p+s$-wave superconductor[14], N is the Bi$_2$Te$_3$ Dirac metal channel and I the interface barrier (see Fig. 1d). The induced gap in S' will define the maximum $I_C R_N$ product where $I_C$ represents the maximum Josephson current and $R_N$ the normal resistance of the junction. For the IVC of Fig. 1c we get an $I_C R_N \approx 100$ μV indicating an induced gap close to that of the Al[22]. This is further supported by the high value of the transparency $D$ we extract from the IVC.

Specifically from the value $eI_{exc}R_N/\Delta_{S'}$, with $\Delta_{S'} \approx \Delta_{Al}$, we can extrapolate the value of the barrier transparency $D \cong 0.8$ [21].

Figure 2 shows the magnetic field dependence of the Josephson current for the same junction as in Fig. 1c. We observe an ideal Fraunhofer pattern with clear secondary lobes (cyan circles). From the modulation period $\Delta B = \phi_0/A_{eff} \approx 2.5$ mT, where $\phi_0$ is the superconducting flux quantum, we extract an effective area $A_{eff}$ of 0.75 $\mu$m$^2$. This value is very close (within few %) to the numerically calculated effective area, by assuming a $2\pi$ periodic current phase relation (see Supplementary Note 1). This has been verified for several junctions on various chips, which confirms a conventional magnetic field behavior of the devices in agreement with the theoretical predictions[23] and various experimental reports[24], while in contrast with the anomalies stated in earlier works[25].

After thermal cycling (from base temperature to 300 K and back to base temperature) the transport properties of all devices undergo a dramatic change. The critical current of the junctions is, for most devices, reduced by more than two orders of magnitudes (see Fig. 2) while the normal resistance increased by a factor between 1.1 and 6 (see Table 1).

We shall argue that this phenomenology is related to the dramatic role played by strain to tune the interquintuple layers interaction and therefore the topological phase in TIs. First principle calculations[26,27] have shown that the consequence of a tensile strain (out of plane compression) is a shift of the Dirac point closer to the valence band while a compressive strain (out of plane expansion) leads to a gap opening at the Dirac point[26]. Strain is usually generated during the epitaxial growth of the material on the substrate, with lattice parameters different from those of the topological insulator. Recent reports have shown the

tunability of the Dirac point by strain in thin topological crystalline insulator SnTe[28] and at grain boundaries in $Bi_2Se_3$ thin films[29].

Our experiment is quite different: exfoliated $Bi_2Te_3$ flakes are transferred to a $SiO_2$/Si substrate, so a possible strain-related phenomenology cannot be attributed to the growth process. The strain instead is related to the huge difference in the thermal expansion coefficient of $Bi_2Te_3$ ($\sim$13.4 × 10$^{-6}$ °C$^{-1}$) and that of the $SiO_2$/Si substrate (0.5 × 10$^{-6}$ °C$^{-1}$/2.4 × 10$^{-6}$ °C$^{-1}$). The flake will experience this difference by the clamping to the substrate, through the patterning of the Al electrodes forming a nanometer sized gap junction. During the warming up of the sample the $Bi_2Te_3$ flake at the nanogap undergoes a compressive strain inducing plastic deformation that leads to a buckling of the $Bi_2Te_3$ channel forming the nanogap upon a subsequent cool down[30]. Figure 3 shows a SEM picture of a junction after cycling; a typical buckling feature appears in the nanogap (see Supplementary Note 3).

Plastic deformations at the nanochannel can substantially affect the Josephson properties. The opening of a gap at the Dirac node, which can eventually reach the bulk gap[26], works as a tunneling barrier reducing the Josephson current as we observe in our experiment. The physics behind this reduction is that the opening of a gap at the Dirac node causes current-carrying surface states beneath the Al electrodes to become evanescent waves in the TI nanogap region, which thus decay over a shorter distance than in the gapless case and thus reduces the critical current.

For the devices, which undergo the most dramatic changes in the Josephson current, the value of the $I_{exc}$ is zero, signifying a very low transparency. However for these samples the most striking experimental observation is the inverted Josephson magnetic field pattern that

appears after thermal cycling (Fig. 2 (blue points) and Fig. 4a, see also Supplementary Figure 2, and Supplementary Note 2).

## Discussion:

Our results are consistent with a proximity-induced superconductivity in the TI surface states compatible with a $p+s$ OP which contains a chiral $p$-wave term. Let's now focus on the effect of the interface transparency on the chiral $p$-wave part of the $p+s$ OP. Self-consistent calculations have shown that the transparency of the interface can have a dramatic effect on a chiral $p_x$+i$p_y$ wave OP. For highly transparent interfaces, both the $p_x$ and $p_y$ component of OP gap are equally suppressed at the interface[31] (see Fig. 5a); however the net pairing symmetry of the order parameter remains a chiral $p$-wave at the interface region, with a slightly reduced magnitude. As the interface transparency is lowered, the $p_x$ component of the order parameter will be strongly suppressed, while the $p_y$ slightly enhanced[31,32] (Fig. 5b). Considering our data at the first cool down the transparency is high; if the induced order parameter is of the type $p+s$ it will preserve the same symmetry at the interface. In this case, a Fraunhofer-like magnetic pattern is expected[23], as we have found in our experiment. After thermal cycling, however, the critical current is suppressed by orders of magnitude (Fig. 2). This fact cannot solely be attributed to a change in interface transparency due to plastic deformations induced by compressive strain (buckling). The strong suppression of $I_C$ is, on the other hand, fully consistent with a change in pairing symmetry of the $p+s$ OP. In particular, as theoretically predicted[31,32] the chiral $p$-wave term will assume a predominantly $p_y$-wave form at the interfaces (with the $p_y$ lobe parallel to the interface).

If one considers only the chiral term of the $p+s$ OP its amplitude and phase at the S'IN interface are shown in Figures 5c and 5d for a highly and low transparent interface

respectively[32]. The chiral $p$ term, of the total OP, is mainly transformed in $p_y$ for low transparency barriers. Theory shows that in $p_y$-$p_y$ junctions the critical current becomes strongly suppressed even for a slight increase of scattering in the junction[33]. This is therefore fully consistent with the observed reduction, of orders of magnitude in $I_C$, after thermal cycling. Secondly, such a change in pairing symmetry also explains the inversion of the Fraunhofer pattern with a dip at $B = 0$ when the S'IN interface transparency is lowered. For a predominantly $p_y$-wave OP at the interface, the occurrence of scattering in the normal region of the junction will couple quasiparticle trajectories, emerging from positive $p_y$ lobe of the OP on one side of the junction, with trajectories that probe the negative $p_y$ lobe on the other side of the junction. This corresponds to a net π-phase shift in the Josephson coupling (see Fig. 6a and 6b). Here it is worth pointing out that while the buckling wave in the nanogap area can partially be responsible for the reduction of the Josephson current (and possibly for a non homogenous distribution of $J_C$ across the width of the junction) it cannot account for the peculiar dip of the Fraunhofer patter at $B$=0, requiring a net π-phase shift between the electrodes. While the above scenario provides a possible route for introducing 0- and π-trajectories in the system, we underline that an s-wave component of the OP simultaneously exists. We speculate that for π-shifted trajectories to become possible, it might be of importance that the $p_y$-wave component is enhanced compared to the $s$-wave one since the $p$- and $s$-wave condensates are not separated due to the spin-mixing resulting from the normal-state TI dispersion with spin momentum locking.

In films with pronounced pyramidal nano-domains, like the flakes used in our experiment, the edges have proven to host an enhanced density of states[34,35] and to be preferential trajectories[36]. Aharonov-Bohm oscillations in the magneto-resistance of thin films, where

orbits around the triangle edges define the loop areas[36] have been recently detected, demonstrating that the triangle's corners and more in general step-edges work as scattering centers. To show that this effect is also observed in our $Bi_2Te_3$ flakes we have measured the magneto-transport down to 20 mK temperatures. We observe clear periodic magneto-resistance oscillations, arising from coherent scattering of electron waves from the corners of the pyramidal domains and/or from the points where two domains merge, with a period corresponding to the morphology of the flake (see Supplementary Note 4). In this scenario, among the quasiparticles crossing the $Bi_2Te_3$ normal channel, the ones undergoing large scattering angles at the corners or step-edges of the pyramidal nano-domain, will probe a phase shift of $\pi$ on the other side of the junction. The quasi-particles trajectories, instead, that do not undergo scattering events in the normal channel (e.g. coupling lobes of the same sign with each other) give a conventional 0 coupling. The schematics of the scattering processes are shown in Fig. 6b. The net result is that the total supercurrent is partially cancelled by these competing contributions, which amounts to a suppressed $I_C$ at zero field.

This simple explanation has also a direct microscopic analog in terms of Andreev levels mediated transport in a Josephson junction[37]. Indeed a peculiarity of our measurement is the transition of the magnetic pattern to a conventional Fraunhofer type, for temperatures around 100 mK when the dip at zero field flattens out (Fig. 4b). This crossover has been theoretically predicted in the framework of low transparent Josephson junctions involving d-wave HTS[37,38]. The physics behind this phenomenology is connected to the presence of zero energy Andreev bound states, or Mid Gap States (MGS), characteristic of Josephson junctions with an OP with a nontrivial internal phase[37]. For the case of Figure 5b, the $p_x$ component, responsible of the MGS formation[31], is non zero on distances where Andreev

reflection takes place[37] (which are given by the coherence length in S'). Midgap states are therefore formed on both sides of the junction and their resonant coupling gives an additional Josephson transport channel proportional to $D^{0.5}$ (instead of $D$ as for conventional Andreev level mediated transport). When the MGS carrying current dominates at very low temperature[37,38] and scattering events take place at triangle corners, the equilibrium phase difference of the system is at $\pi$, corresponding to a negative current. For straight quasiparticle trajectories the MGS current gives a conventional 0 phase character ground state.

Formally, this scenario is similar to artificially designed 0-$\pi$ ferromagnetic Josephson junctions where multiple connected segments are $\pi$-shifted compared to each other[39]. In this case, a similar inverted pattern is observed. The dip of the Josephson current at $B = 0$ is thus the proof of a superconducting order parameter with an internal phase-structure, which can cause either 0 or $\pi$ couplings, precisely as seen in our setup (see Supplementary Note 5).

In a generic junction, the contribution to the current is due to both the resonant MGS and the conventional Andreev bound states. At larger temperature the contributions of MGSs current cancels out since both MGSs carrying forward and backwards current are populated[37] and therefore the conventional Andreev levels current will dominate. This leads to a peculiar $\pi$ to 0 transition by increasing the temperature. Signatures of such a crossover have been reported in a Superconducting Quantum Interference Device (SQUID) geometry for $d$-wave junctions[40]. However those results were never confirmed, posing serious doubts on the effective possibility to detect a $\pi$-0 transition from the temperature dependence of the

magnetic patterns. Our experiment is therefore one of the first demonstrations of Josephson transport mediated by MGS.

Clearly the dip at $B$=0 in the magnetic pattern is also compatible with a $d_{x^2-y^2}$ wave order parameter[41]. We can rule out this possibility since the $d_{x^2-y^2}$ OP does not change symmetry depending on the interface transparency[37] and therefore cannot account for the dramatic modifications of the magnetic field response of the junctions upon thermal cycling. An helical $p_x+p_y$ OP can instead be compatible with our entire experimental scenario. However this OP symmetry does not come directly from the theoretical works[1,14,41] so the possibility to induce an helical $p_x+p_y$ OP needs further theoretical assessment.

To further check the conjecture that step-edges and pyramidal corners act as scattering centers, we have fitted the temperature dependence of the critical current $I_C(T)$ of the high transparency junctions. We consider that our S'INIS' junction is characterized by insulating barriers with transparencies $D_1$ and $D_2$, which are separated by a distance $L$ [42]. The disorder in the normal channel is characterized by a mean free path $l_e$. As previously discussed, in case of high transparency barriers the junction behaves similarly to conventional $s$-wave junctions[43]; in this case we do not lose generality by modeling the S' superconductor with an $s$-wave OP. The Josephson current for an S'INIS' junction in the clean limit $l_e \gg L$ has been derived in a previous work[8]. To fit our data, we have modified the original result of Ref. [8] in two ways (see Supplementary Note 6): (i) by adapting it to the two dimensional Dirac character of the surface states, and (ii) by extending it into the regime $l_e \approx L$ which is achieved by changing the flight time expression of the electrons between the leads. We have verified that our approach well reproduces the Eilenberger theory for purely

ballistic junctions and the Usadel theory for short diffusive junctions. Figure 7 shows the $I_C(T)$ of the junction of Fig. 1c. The solid curve is the fit by considering the geometrical dimensions and $l_e = 130$ nm. The extracted value for the mean free path clearly supports our assumption that few scattering centers are present in the barrier and that some quasiparticles while crossing the TI channel will undergo a single scattering event. The value of the normal state resistance extracted from the fit is higher than the experimental one, which can be explained by shunting effect of the bulk of the $Bi_2Te_3$[8] (see Supplementary Note 6).

It is worth mentioning that we did not observe any apparent correlation between the orientation of the nano-pyramidal domains and that of the electrodes. This is not surprising; the basic mechanism to get an inverted magnetic field pattern, is the occurrence of 0 and $\pi$ trajectories within the nanogap and they can be obtained for whatever orientation of the triangles with respect of the electrode, provided that some of the corners and/or the points where two domains merge fall within the $Bi_2Te_3$ nanogap (see Supplementary Note 4).

Finally we would like to point out that magneto transport measurements (Supplementary Note 4) have clearly shown that scattering mechanisms, connected to the morphology of the $Bi_2Te_3$ flakes, take place in the junction channel, which is instrumental to get $\pi$ trajectories. However the interpretation of our measurements does not change if other scattering mechanisms are also involved.

To conclude it is worth discussing if the interpretation we have provided of our experiment would change if we would also consider a parallel Josephson contribution due to the transport through the bulk with an *s*-wave OP. Cooper pair transport through the bulk

would just act as 0 trajectories effectively lifting the value of the Josephson current at zero external magnetic field, however leaving unaltered the dip structure at zero field of the magnetic pattern.

# Methods:

## Materials and device fabrication

$Bi_2Te_3$ thin films have been deposited by molecular beam epitaxy on a GaAs (100) substrate with a 2° vicinal cut. Prior to the growth, a one minute Te soaking to the GaAs surface was carried out for passivation purpose. Then, the growth started at a temperature of 180 °C, when both Te and Bi sources were opened, and ended at a thickness of 80 nm after one hour. The films have been grown with a high Te/Bi flux ratio, to obtain high quality crystallinity, as verified by X-Ray Diffraction (XRD) analysis[44].

The films show characteristic aligned pyramidal domains (see Fig. 1b). This morphology is different with respect to the growth on exact substrates, where two types of pyramidal domains rotated by 60° with respect to each other are observed[20]. As verified by XRD analysis, our films show 86% single domain, which indicates that the vicinal cut in the substrate leads to suppression of twin boundaries, as also reported in literature[20]. The transport properties have been characterized by magneto-transport measurements; the films show an effective surface carrier density of $4\times10^{13}$ $cm^{-2}$ and mobility above 2000 $cm^2V^{-1}s^{-1}$ (at the temperature of T = 2 K).

Josephson junctions using Al electrodes have been realized by transferring the flakes on a $Si/SiO_2$ substrate. To improve the adhesion with the Al electrodes a thin layer (3 nm) of Pt is

deposited in situ on the surface of the flake, previously cleaned by a short (10 s) Ar⁺ ion milling[45,46].

**Data availability**

All relevant data are available from the authors.

## Acknowledgments:

## Author contributions:

S.C. and L.G. have fabricated the samples and performed the measurements. S.W. and Y.S have grown the Bi2Te3 thin films. D.G. and J.L. have contributed with theoretical insights and F.T. with discussions. G.K., E.O., A.K., R.A., R.B. have performed morphological and structural characterizations of the devices. F.L. and T.B. have interpreted the measurements in collaboration with all authors. F.L has written the paper with inputs from all co-workers.

## Additional information:

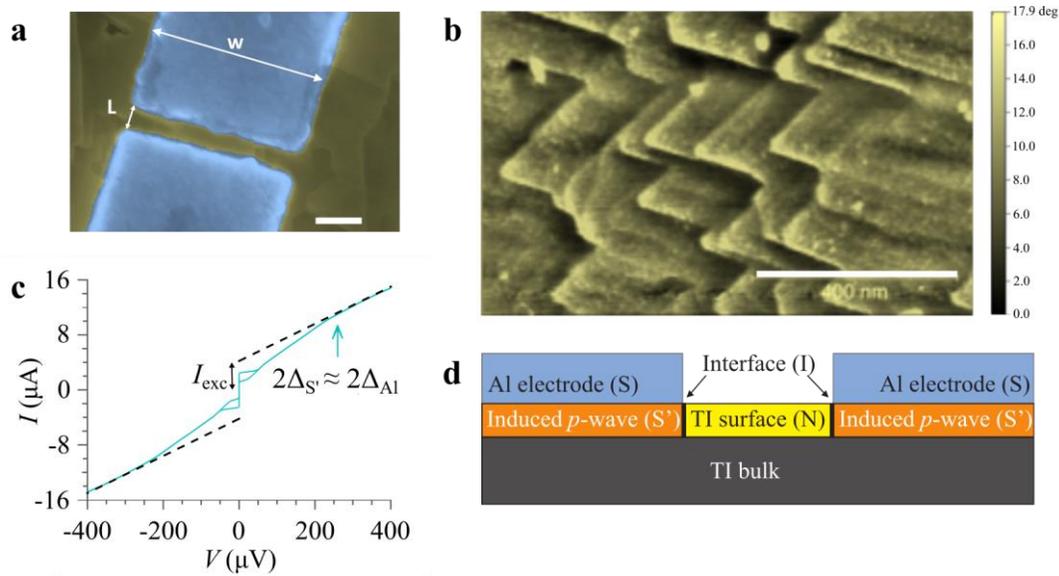

**Fig. 1 Morphological and transport characterization of a typical proximity based Bi₂Te₃ Josephson junction. (a)** Scanning Electron Microscope image of a junction made using an exfoliated flake. The light blue areas define the Al/Pt electrodes while the Bi₂Te₃ flake is shown in yellow. Scale bar: 200 nm. **(b)** Atomic Force Microscope image of a Bi₂Te₃ thin film grown on GaAs substrate showing a typical growth with aligned pyramidal domains. Scale bar: 400 nm. **(c)** IVC of a typical junction ($w = 1$ μm, $L = 200$ nm) measured at 20 mK (cyan). The resistive branch shows a neat change of slope at a voltage $V \approx 260$ μV (cyan arrow) that corresponds to approximately $2\Delta_{Al}$ for the thickness of the electrodes[22]. The black arrow indicates the value of the extracted excess current. **(d)** Cross section schematics of the effective device under consideration. The transport properties can be assimilated to those of a S'INIS' Josephson junction where S' represents the induced superconductor in the Bi₂Te₃.

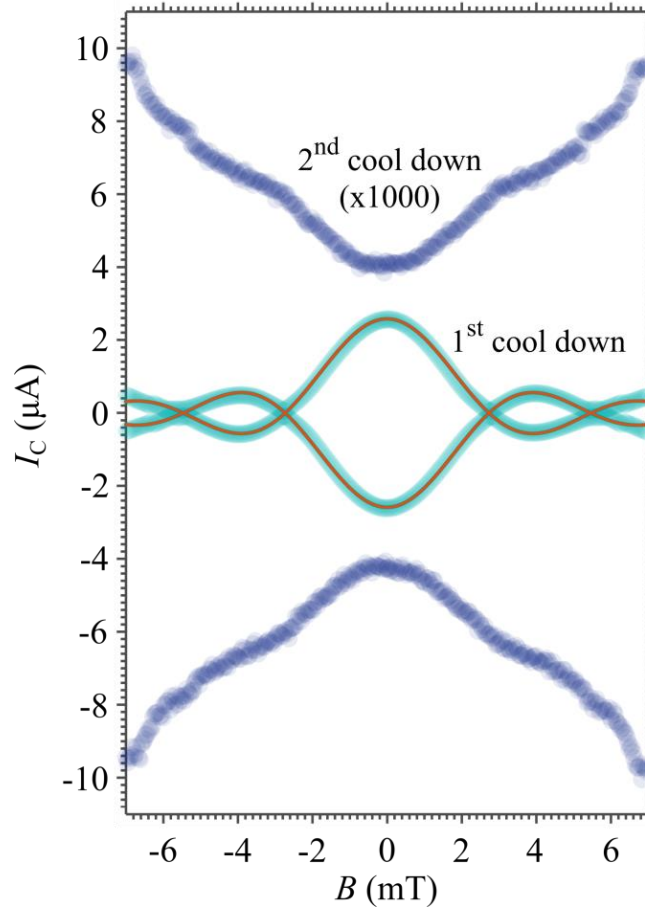

**Fig. 2 Magnetic field pattern of a typical junction before and after the thermal cycle.**
Critical current $I_C$ dependence on the externally applied magnetic field $B$ for the junction of
Fig. 1c at 20 mK, before (cyan) and after (blue) the thermal cycle. At the first cool down the
$I_C(B)$ dependence shows a conventional Fraunhofer dependence fitted with the expression
$I_C(B) = I_C(0) \left| \frac{\sin(\pi \phi / \Phi_0)}{(\pi \phi / \Phi_0)} \right|$ (red line) where $I_C(0)$ is the critical current at zero
field, $\phi_0$ is the flux quantum, and $\phi = BA_{eff}$ is the flux through the junction, being $A_{eff}$ the
effective area. After the thermal cycle (blue points), the critical current is dramatically
reduced (the data points are multiplied by a factor 1000 for clarity), and a dip at $B = 0$
appears.

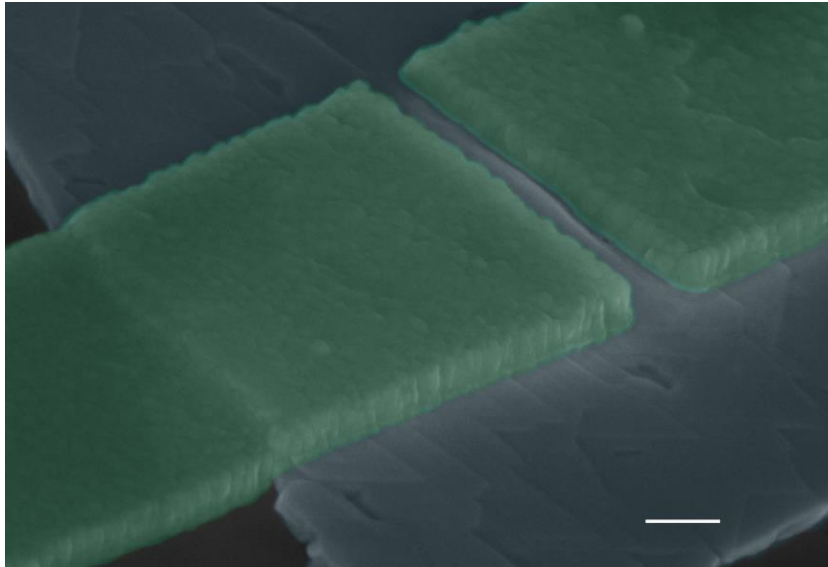

**Fig. 3 Colored SEM image of a typical junction after the second cool down.** The picture (scale bar: 200 nm) show a junction presenting an inverted Josephson magnetic pattern, after the second cool down. A clear buckling feature, induced by a compressive strain, is visible in the nanogap.

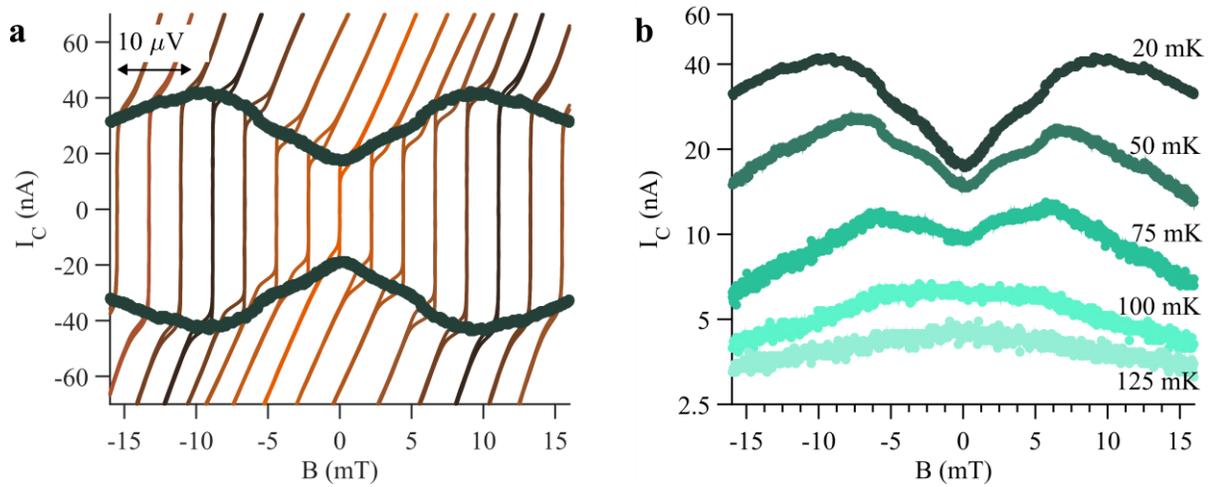

**Fig. 4 Current voltage characteristics as a function of magnetic field and temperature for a typical junction after thermal cycling. (a)** IVCs of a second device ($w = 500$ nm and $L = 100$ nm) are shown as a function of $B$ after thermal cycling. The curves are shifted in voltage by a value proportional to the magnetic field. The green points represent the magnetic pattern of the junctions obtained using a voltage criterium $V = 600$ nV. Also in this case the magnetic pattern presents a very pronounced dip at $B = 0$. **(b)** Evolution of the magnetic pattern shown in panel (a) as a function of the temperature. The dip flattens out at a temperature close to 100 mK.

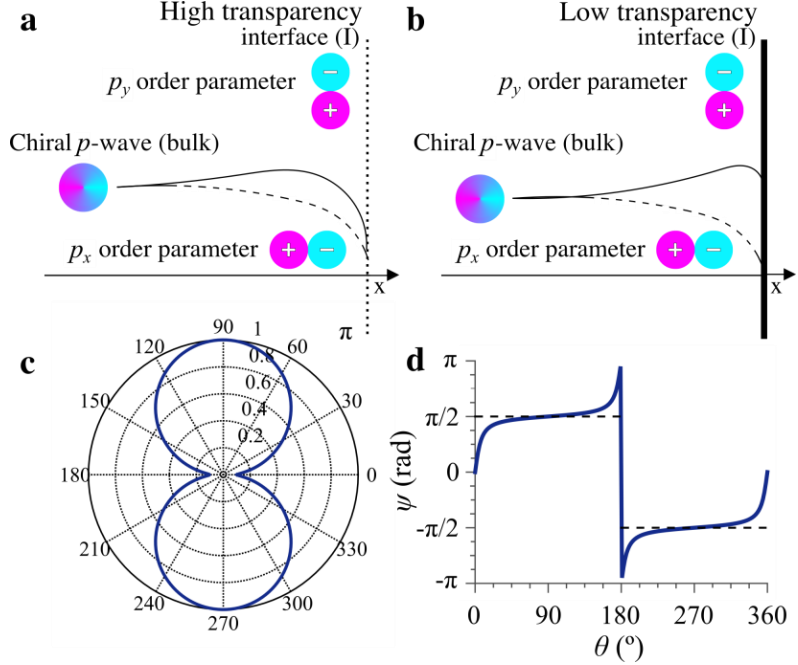

**Fig. 5 Proximity induced OP on the surface of Bi₂Te₃.** We consider only the chiral $p_x$+i$p_y$ part of the total $p+s$ OP. Close to the interface I the chiral term is modified: the $p_x$ (dashed line) and $p_y$ (straight line) components exhibit different magnitude. In case of high transparency interface **(a)** the two components are both suppressed in equal manner and the resulting interface OP still preserve the $p_x$ + i$p_y$ wave character. For a low transparency interface **(b)**, the $p_x$ component is highly suppressed while the $p_y$ slightly enhanced. The resulting OP has a predominantly $p_y$ character. Panel **(c)** shows the amplitude of the resulting interface OP by considering a mixture $p_y$+i$\alpha p_x$, with $\alpha$ = 0.1. This corresponds to the reduction of $p_x$ on distances of the order of the coherence length of S' [28]. The corresponding phase is shown in **(d)**.

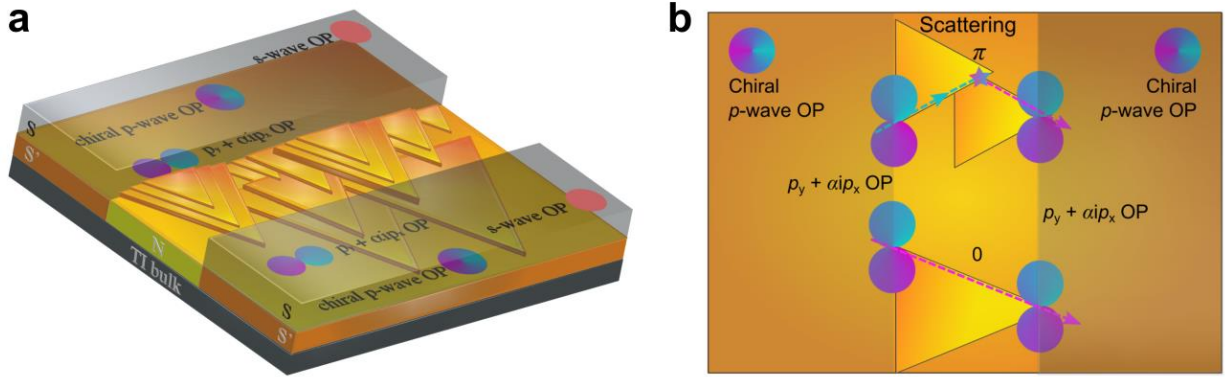

**Fig. 6 Sketch of the device showing the induced order parameter and the occurrence of π-paths across the junction. (a)** Three dimensional sketch of the Bi₂Te₃ device. The Al electrodes (S) induce an effective *p+s*-wave superconductivity on the surface of the TI. In the picture we consider only the chiral term of the total OP. In close proximity of a low transparency interface the chiral $p_x$+i$p_y$ changes symmetry resulting in a predominant $p_y$ component (see main text for details). **(b)** Top view of the device. The occurrence of scattering at the pyramid corners and/or at the merging points of two pyramidal domains, in the N region of the junction, makes quasiparticle trajectories emerging from a negative $p_y$ lobe (blue arrow) on one side of the junction to couple with trajectories that probe the positive $p_y$ lobe (red arrow) on the other side of the junction. In case of scattering-free transport in the N region the quasiparticles trajectories probe the same phase in both electrodes (red arrows).

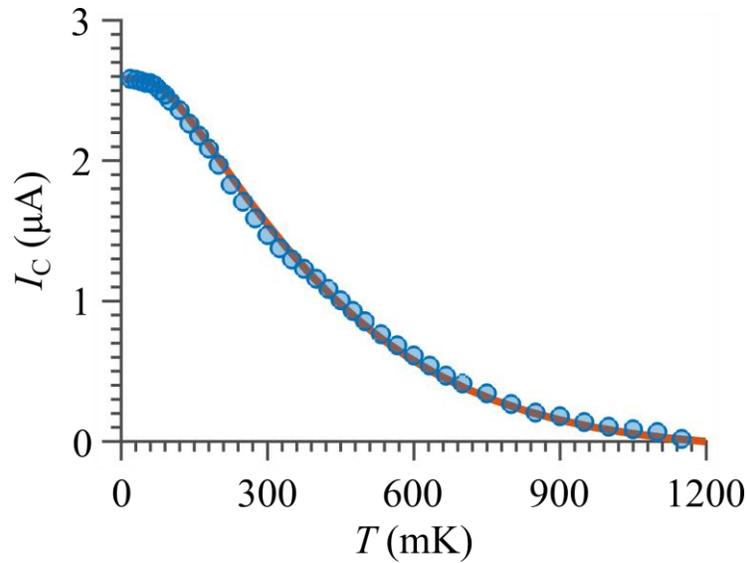

**Fig. 7 Temperature dependence of the Josephson current for a high transparency junction.** Critical current as a function of the temperature, measured at the first cool down for the junction of Fig. 1c (full circles). The red line is the fitting considering a quasi-ballistic transport model (see Supplementary Note 6 and Supplementary Equations (5-10)). We extract the following fitting parameters for the mean free path $l_e = 130$ nm and the transparencies of the two barriers $D_1 = D_2 = 0.98$. From Supplementary Equation (13) the expected normal state resistance is $R_N = 70$ Ω. The experimental value for $R_N$ is instead 35 Ω. The discrepancy between these two values can be attributed to the shunting effect of the bulk of the Bi$_2$Te$_3$ flake.

| Device | $I_\mathrm{C}^{\text{before}}$ | $I_\mathrm{C}^{\text{after}}$ | $R_\mathrm{N}^{\text{before}}$ | $R_\mathrm{N}^{\text{after}}$ | $I_\mathrm{C}R_\mathrm{N}^{\text{before}}$ | $I_\mathrm{C}R_\mathrm{N}^{\text{after}}$ | $w$ | $L$ | $\dfrac{I_\mathrm{C}^{\text{after}}}{I_\mathrm{C}^{\text{before}}}$ | $\dfrac{R_\mathrm{N}^{\text{after}}}{R_\mathrm{N}^{\text{before}}}$ |
|---|---|---|---|---|---|---|---|---|---|---|
| | μA | μA | Ω | Ω | μeV | μeV | μm | nm | | |
| **JM10** | 2.6 | 0.002 | 36 | 215 | 89 | 0.4 | 1 | 200 | 0.07% | 6.0 |
| **JM5** | - | 0.017 | - | 216 | - | 3.6 | 0.5 | 100 | - | - |
| **JM11** | 1.8 | 0.085 | 41 | 62 | 47 | 5 | 1 | 300 | 4% | 1.5 |
| **SF2** | 0.58 | 0 | 15 | 20 | 9 | 0 | 2 | 100 | 0% | 1.3 |
| **JM1** | 0.6 | - | 86 | - | 54 | - | 0.5 | 200 | - | - |
| **SM4** | 9.3 | 0.060 | 17 | 61 | 144 | 8.6 | 1.5 | 100 | 0.6% | 3.6 |
| **SD3** | 0.36 | 0.040 | 18 | 53 | 6 | 2 | 2 | 250 | 11% | 2.9 |
| **SF3** | 2.2 | 0.030 | 10 | 25 | 22 | 0.7 | 2 | 100 | 1.4% | 2.5 |
| **SF8** | 3.2 | 0.180 | 11 | 25 | 35 | 4.5 | 2 | 100 | 5.6% | 2.3 |
| **SD4** | 0.6 | 0.02 | 32 | 52 | 2 | 1 | 2 | 250 | 3% | 1.6 |
| **SF4** | 19.8 | 6.600 | 3.6 | 5.6 | - | - | 6 | 100 | 33% | 1.5 |
| **SD2** | 0.52 | 0.012 | 17 | 20 | 9 | 0.2 | 4 | 250 | 2.3% | 1.2 |
| **SD1** | 1.2 | 0.086 | 27 | 30 | 32 | 2.5 | 2 | 250 | 7.1% | 1.1 |
| **SM1** | 11.0 | - | 15 | - | 151 | - | 1.5 | 100 | - | - |

**Table 1: Relevant parameters for all the measured devices before and after the thermal cycle.**
The critical current $I_\mathrm{C}$, the normal resistance (measured above gap) $R_\mathrm{N}$, the $I_\mathrm{C}\,R_\mathrm{N}$ product, the width of the device $w$ (in case of a SQUID twice the junction) and the distance between the electrodes $L$ are reported. Devices JM10, JM5, SM4, SF8 and SD1 showed inverted magnetic patterns in the second cool down. The devices JM10 and JM5 are those, whose data are shown in the text. The junctions start with a J in the device name, whereas the SQUIDs device names start with an S.



\fancyhf{}
\fancyhead[L]{ščđćž}

\beginsupplement

\section{Effective area of a planar Josephson junction}

\begin{figure}[!ht]
\includegraphics[width=8cm]{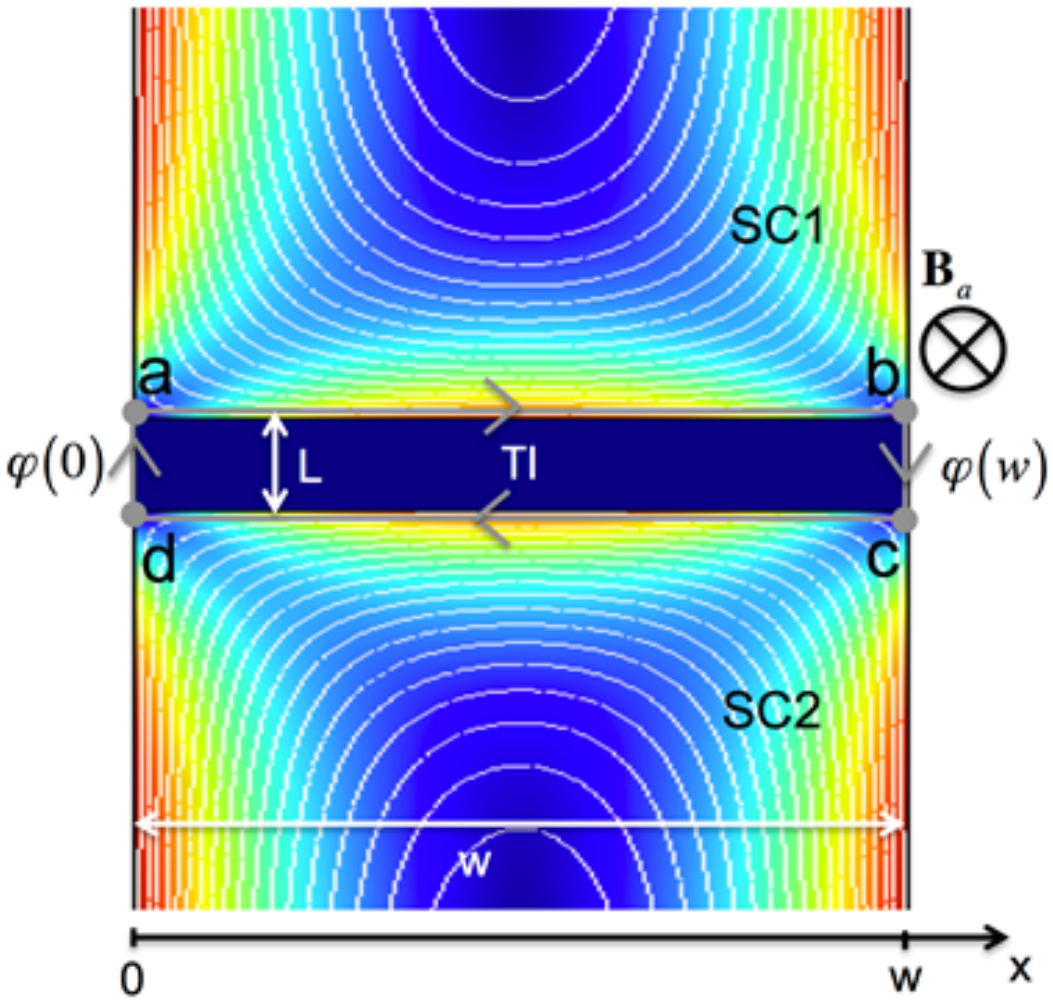}
\caption{\textbf{Supercurrent density distribution in a planar Josephson junction.} Sketch of two superconducting electrodes (SC1, SC2) contacted to a normal conducting channel (TI). The gray loop indicated the path integral used in Supplementary Equation (4). $B_\mathrm{a}$ is the externally applied magnetic field. The white lines indicate the current flow direction of the screening current and the color-coding in the electrodes reflects the amplitude of the screening current density (red: large, blue: small).}
\label{fig:FigS1}
\end{figure}

The effective area $A_{\mathrm{eff}}$ of a planar junction in the thin film limit $t < 2\lambda$, with $t$ the film thickness and $\lambda$ the London penetration depth, having a junction width, $w$, much smaller than the Pearl length, $\lambda_\perp=\lambda^2/t$, is well approximated by the following expression\cite{rosenthal1991flux}
\begin{eqnarray}
A_{\mathrm{eff}} \approx wL+w^2/1.82 \;.
\label{eq:eq1}
\end{eqnarray}
Here $L$ is the distance between the electrodes (see \ref{fig:FigS1}). For junction widths larger than the Pearl length, Supplementary Equation (\ref{eq:eq1}) is not anymore valid and the effective area has to be computed numerically.

The effective area can be determined numerically as it follows.

First we calculate the Meissner screening currents in the superconducting electrodes by solving the Maxwell London equations on the junction geometry in the presence of an externally applied magnetic field:
\begin{eqnarray}
\mu_0 {\boldsymbol \nabla} \times (\lambda^2 \cdot \bm{\mathrm{j}}) + \bm{\mathrm{B}} = 0  \; ; \;\;\;\;\;\;\;\;\;\; {\boldsymbol \nabla} \times \bm{\mathrm{B}} = \mu_0\bm{\mathrm{J}} \; .
\label{eq:eq2}
\end{eqnarray}
Here $\mu_0$ is the vacuum permeability, $\bm{\mathrm{j}}$ is the supercurrent density in the electrodes, and $\bm{\mathrm{B}}$ is the magnetic induction containing the externally applied field, $B_\mathrm{a}$, and the field generated by the supercurrents. 

Next we need to calculate the difference between the gauge invariant phase differences across the junction at the left edge $\phi(0)$ and right edge $\phi(w)$ of the junction (see \ref{fig:FigS1}). From the above calculated Meissner currents and resulting magnetic fields we can readily compute $\Delta \phi = \phi(w) - \phi(0)$ making use of the property that the superconducting phase is single valued in the electrodes:
\begin{eqnarray}
\Delta\phi = \frac{2\pi}{\phi_0} \oint \bm{\mathrm{A}} \cdot \partial \bm{\mathrm{l}} + \mu_0 \lambda^2 {\left[\int_{a}^{b} \bm{\mathrm{j}} \cdot \partial\bm{\mathrm{l}} + \int_{c}^{d} \bm{\mathrm{j}} \cdot \partial\bm{\mathrm{l}} \, \right]} \; .
\label{eq:eq3}
\end{eqnarray}
Here $\phi_0 = 2 \cdot 10^{-15}$ Vs is the superconducting flux quantum and $\bm{\mathrm{A}}$ is the vector potential with $\bm{\mathrm{B}} = {\boldsymbol \nabla} \times \bm{\mathrm{A}}$. The closed path integral in the first term of the right hand side of the equation is taken along a loop enclosing the two edges of the junction at $x=0$ and  $x=w$. For the loop indicated in \ref{fig:FigS1} the closed path integral corresponds to the total magnetic flux through the normal conductor in the junction. The line integrals in the second term of the right hand side of the Supplementary Equation (\ref{eq:eq3}) are taken only within the superconducting electrodes, i.e. from point $a$ to $b$ in the upper electrode, SC1, and from point $c$ to $d$ in the lower electrode, SC2 (see \ref{fig:FigS1}).

The effective area of the junction is finally obtained using the following standard expression:
\begin{eqnarray}
A_{\mathrm{eff}} = \frac{\phi_0}{2\pi} \frac{\phi(w) - \phi(0)}{B_\mathrm{a}} \; .
\label{eq:eq4}
\end{eqnarray}
In the above equation we assume a $2\pi$ periodic current phase relation along the junction.

For typical values of the London penetration depth ($\lambda \approx 100$ nm) for our 90 nm thick Al films\cite{romijn1982critical}, we obtain effective areas in the range $0.68-0.70$ $\mu$m$^2$, which is very close to the experimental value of $0.75 \, \mu$m$^2$. This strongly supports that a $2\pi$ periodic current phase relation mainly dominates the magnetic pattern of the critical current in our junctions.

\section{Effects of thermal cycling and occurrence of $0-\pi$ transitions in the magnetic pattern}

We have measured a total of 14 devices, including Josephson junctions and dc SQUIDs (see Table 1 in the main text), all fabricated with flakes from the same Bi$_2$Te$_3$ film. 12 devices have been measured in a second cool down; 5 of them showed unconventional magnetic patterns, with a minimum at zero field.

\begin{figure}[!ht]
\includegraphics[width=15cm]{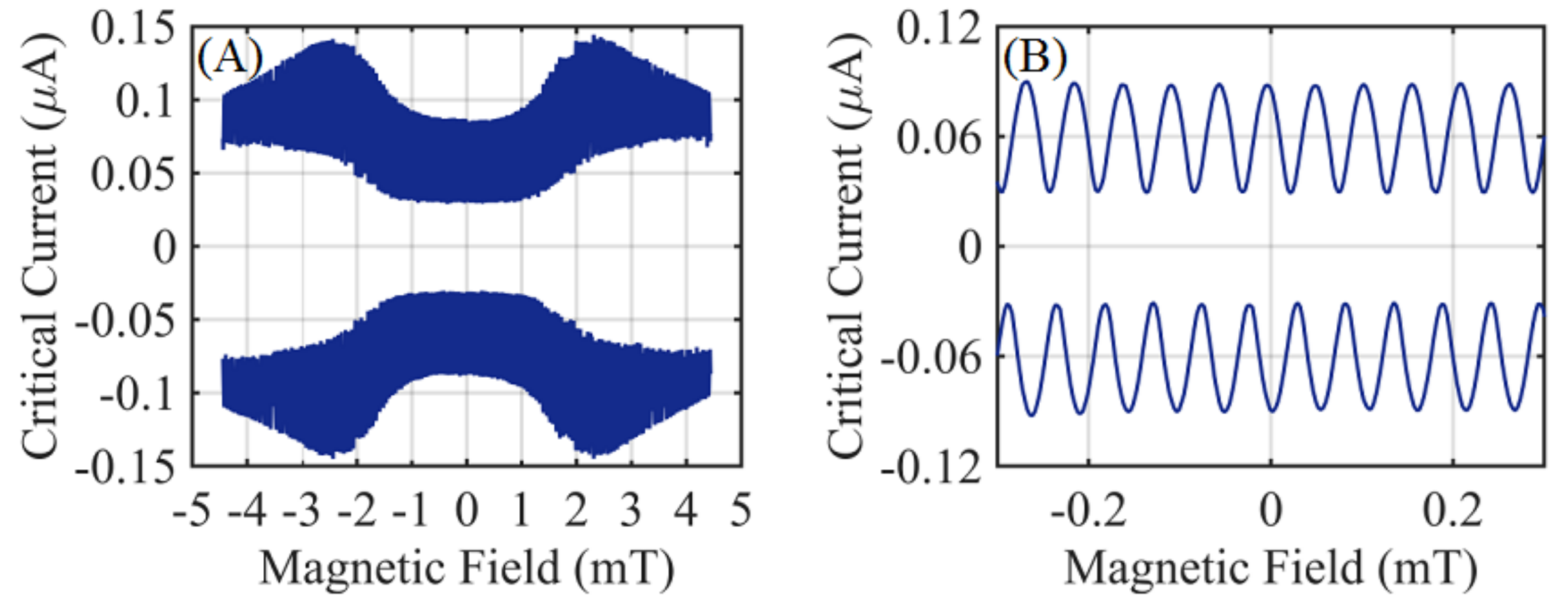}
\caption{\textbf{Critical current dependence as a function of the externally applied magnetic field of the SQUID FS8 at} $\bm{T=20}$ \textbf{mK.} Panel (\textbf{A}) clearly shows the feature of the magnetic response of the single junctions forming the SQUID, while the modulation due to the SQUID loop cannot be resolved on this scale (see text for details). The SQUID modulation becomes visible at low external magnetic fields (panel (\textbf{B})).}
\label{fig:FigS2}
\end{figure}

\ref{fig:FigS2} shows another example of a high field inverted pattern for a dc SQUID (SF8). In this case there are two different modulations of the critical current: one is determined by the SQUID loop and the other is determined by the magnetic response of the single junctions forming the SQUID, which gives a dip at zero magnetic field in the convolution of the maxima of the SQUID modulation (\ref{fig:FigS2}A). \ref{fig:FigS2}B shows the magnetic pattern of the SQUID at low fields.

\section{Microscopic origin of the thermal cycling effect}

The thermal cycling effect has its origin in the role played by strain to tune the topological phase in  TIs.  In our devices, the complex interplay between the  thermal properties of the substrate and the flake, the specifics of the interface  between the Pt sticking layer and the Al,  and the nanoscale dimensions of the  devices have  a fundamental role in  creating plastic deformation at the TI nanogap separating the superconducting electrodes. The irreversible consequences of a compressive strain, induced by thermal cycling, manifest as peculiar buckling waves at the nanogap and a dramatic reduction of the Josephson current. 

First principle calculations\cite{liu2011theoretical, young2011anisotropic} have demonstrated that  an in-plane tensile strain (pulling) leads to a out of plane compression where the quintuple layers are getting closer. An in plane compressive strain (pushing) causes instead an out of plane expansion where the  distance  between quintuple layers is enhanced. The interquintuple interaction plays a dominant role in determining the topological phase: the consequence of an out of plane compression is a shift of the Dirac point closer to the valence band while an out of plane expansion leads to a gap opening at the Dirac point\cite{liu2011theoretical}. This argument is also used to explain why for example Sb$_2$Se$_3$ (with a larger c/a ratio compared to Bi$_2$Se$_3$) that is expected to be a TI in many ways, is instead a trivial insulator. It also tells that, in principle, strain can provide a tuning of the surface states of a topological insulator. 

Strain is usually generated during the epitaxial growth of the material on the substrate, with lattice parameters different from those of the topological insulator.  Recent reports have shown the tunability of the Dirac point with strain in thin topological crystalline insulator SnTe \cite{zeljkovic2015strain} and at grain boundaries in Sb$_2$Se$_3$  thin films\cite{liu2014tuning}.

Our experiment is quite different  since our devices employ Bi$_2$Te$_3$ flakes  exfoliated and transferred to a SiO$_2$/Si substrate, so  a possible  strain-related phenomenology  cannot be attributed to the growth process.

The physics behind the peculiar transport properties of our devices is  instead related to the enormous difference in the thermal expansion coefficient $\alpha$ of  Bi$_2$Te$_3$ ($\approx 13.4 \times 10^{-6}$ $^{\circ}$C$^{-1}$) and that of the SiO$_2$/Si substrate ($0.5 \times 10^{-6}$ $^{\circ}$C$^{-1}$/$2.4 \times 10^{-6}$ $^{\circ}$C$^{-1}$) where the flake is transferred. In our device configuration the Bi$_2$Te$_3$ is not grown on a SiO$_2$/Si substrate: the flake therefore will  experience the huge difference in the thermal expansion coefficient only by the clamping to the substrate, which occurs through the patterning of the Al electrodes forming a nanometer sized gap junction.

At the first cool down, from room temperature to mK, the flake  is prone to contract.  However the Al electrodes will anchor the flake to the substrate making it experiencing the much smaller thermal coefficient of the SiO$_2$/Si. This leads to a tensile  strain (pulling of the flake)  which strongly concentrated  in the part of the flake  located at the nanogap. This important fact is related to the thermal expansion coefficient of Al ($\approx 23 \times 10^{-6}$ $^{\circ}$C$^{-1}$) not much different from that of the Bi$_2$Te$_3$. For a good bonding between the Al and the flake, as the one provided by the Pt sticking layer, one can consider the Al electrodes  and the flake under them as forming a quite homogenous material with a thickness  more than twice  that of the flake (the thickness of Al is roughly twice that of the flake).   In this way most of the strain, provided by the Al electrodes  and locking the flake to the substrate, is located at the thin  Bi$_2$Te$_3$ nanogap,  with a smaller thickness  compared to that of the composite material Al/flake.

Since the tensile strain concentrates in the nanogap region, in this part of the flake the Dirac node is shifted towards the valence band which makes the Bi$_2$Te$_3$ channel effectively more doped.

During the warming up of the sample the Bi$_2$Te$_3$ at the nanogap undergoes instead a compressive strain which opens a gap at the Dirac point. This  compressive strain induces plastic deformation (the sample  properties are completely changed after thermal cycling) and,  if  it overcomes a critical strain, can lead to a buckling of the Bi$_2$Te$_3$ channel forming  the nanogap.  

\begin{figure}[!ht]
\includegraphics[width=14cm]{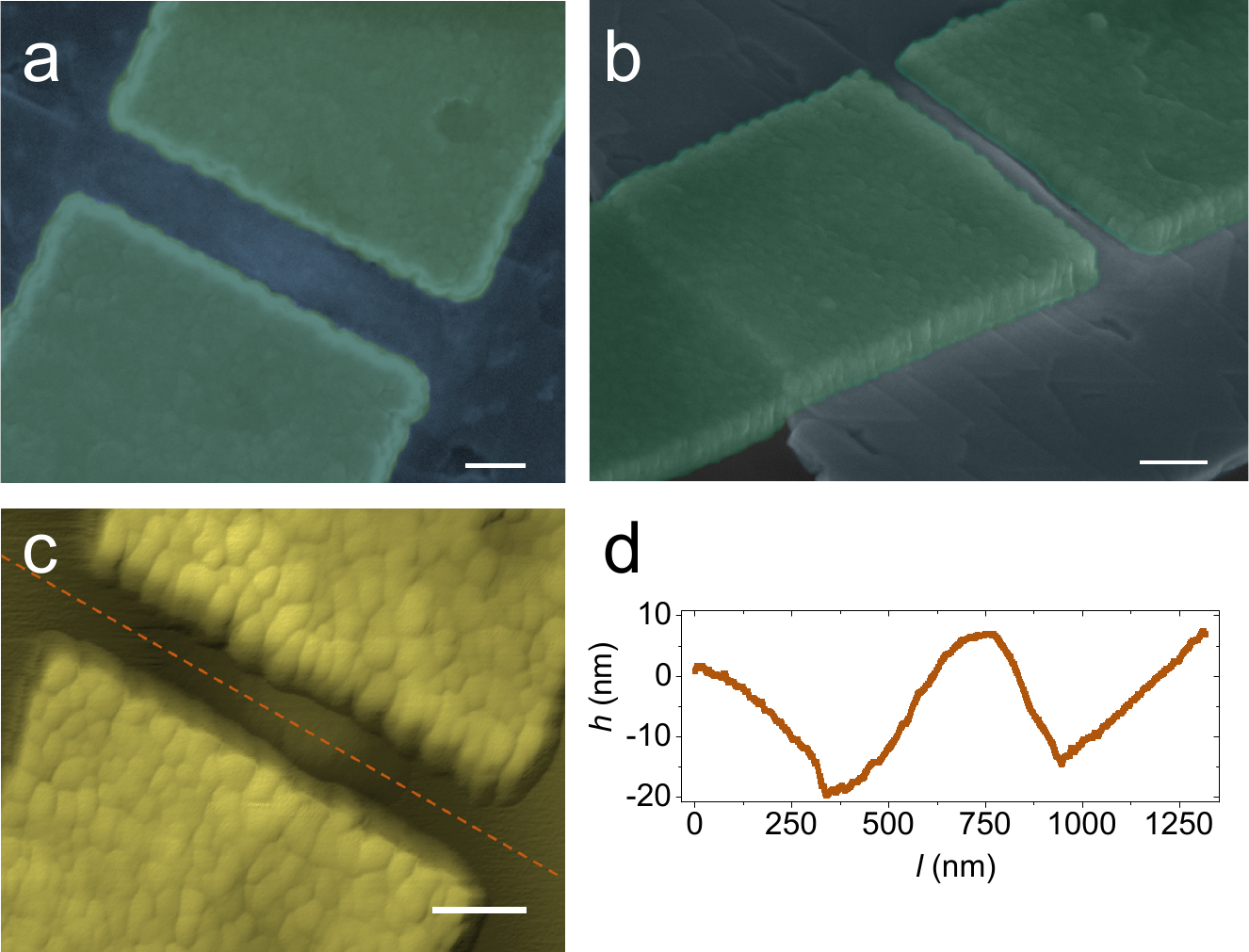}
\caption{\textbf{Colored SEM and AFM images for a junction showing a 0-}$\bm{\pi}$ \textbf{magnetic pattern.} This device is one of those discussed in our main manuscript (Pt interlayer). (\textbf{a}) SEM  top view (scale bar: 200 nm); (\textbf{b}) tilted view (scale bar: 200 nm), green color represents the electrodes; (\textbf{c}) AFM height image (scale bar: 150 nm); (\textbf{d}) height profile along the junction.}
\label{fig:FigS2b}
\end{figure}

\ref{fig:FigS2b}a shows a SEM  top picture of  one of the sample discussed in our paper. 
This specific picture  does not show, very clearly, possible  anomalous features of the flake inside the nanogap,  except a region with a slightly  brighter color. However by taking a SEM picture at an angle, shown in \ref{fig:FigS2b}b, the brightness at the nanogap clearly manifests as a buckling feature\cite{bowden1998spontaneous}. To further evaluate the characteristics of such a buckling, we have performed AFM analysis to resolve the shape of the deformation inside the nanogap. \ref{fig:FigS2b}c shows an AFM picture of the same junction, while \ref{fig:FigS2b}d a line scan (orange dashed line) taken inside the nanogap.  The line scan has a very clear sinusoidal-like shape, a distinctive  feature of a bucking phenomenon induced by compressive strain.

\subsection{Role of the interlayer material: comparison between transport properties of \NoCaseChange{Al/Pt/Bi$_2$Te$_3$/Pt/Al}   and \NoCaseChange{ Al/Ti/Bi$_2$Te$_3$/Ti/Al}}

\begin{figure}[!ht]
\includegraphics[width=8cm]{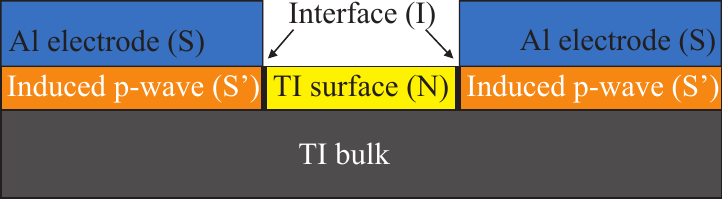}
\caption{\textbf{Cross section schematics of the effective device under consideration.} The transport properties can be assimilated to those of a S'INIS' Josephson junction where S' represent the induced chiral p-wave superconductor in the TI.}
\label{fig:FigS3}
\end{figure}

In our previous works \cite{galletti2014influence, galletti2014josephson},  we have studied the transport properties of  Al/Pt/Bi$_2$Se$_3$/Pt/Al  and Al/Ti/Bi$_2$Se$_3$/Ti/Al   where the flakes were exfoliated from single crystals.  The comparison showed much larger critical current densities and $I_\mathrm{C}R_\mathrm{N}$ product for devices obtained with a Pt interlayer, both facts pointing towards a higher interface transparency barrier (I)  (see \ref{fig:FigS3}), instrumental to observe Majorana bound state 
physics.

We have repeated this experiment on the Bi$_2$Te$_3$ flakes  of this work, with the peculiar triangular morphology,  by using a Ti interlayer. 

\ref{tab2} summarizes the main transport properties of the junctions fabricated using both Pt and Ti. We confirm the general trend: the $J_\mathrm{C}$ and $I_\mathrm{C}R_\mathrm{N}$ values for the Pt interlayers are on average  more than 5 times higher than those obtained with Ti.  The  same trend,    higher values for Pt junctions, is observed for the average transparencies of the barrier I.  This comparison is made at the first cool down.  

\begin{table}[hbpt!]
\resizebox{13.2cm}{!}{
\begin{tabular}{| c | c | c | c |}
\hline
  & $J_\mathrm{C}$ ($\mu$A/$\mu$m)  & $I_\mathrm{C}R_\mathrm{N}$ ($\mu$eV) & $\tau$ (transparency) \\

\hline
  
Rough flakes, \textbf {Pt} interlayer & $2.7 \pm 0.8$ & $79 \pm 19$ & $0.67 \pm 0.05$  \\
\hline
Rough flakes, \textbf {Ti} interlayer & $0.41 \pm 0.14$ & $13 \pm 3$ & $0.49 \pm 0.01$  \\
\hline
\end{tabular}}
\caption{\textbf{Main parameters  extracted from the IVCs of the Bi$_2$Te$_3$ junctions with different interlayer metals}. \label{tab2}} 
\end{table}

\begin{figure}[!ht]
\includegraphics[width=14cm]{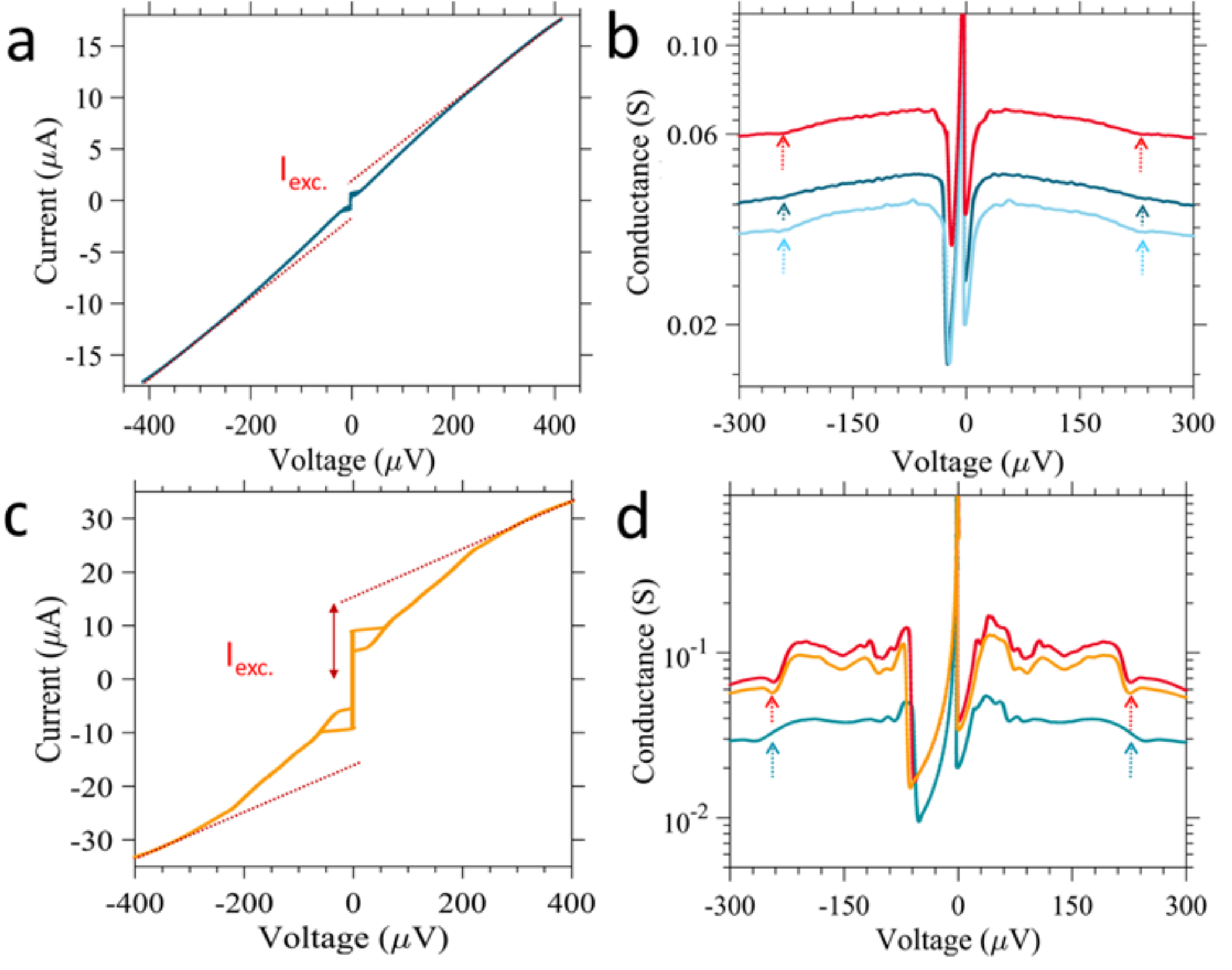}
\caption{\textbf{IVCs and conductance spectra for  different interlayer metals.} IVCs of the Bi$_2$Te$_3$ junctions with (\textbf{a}) - Ti and (\textbf{c}) - Pt interlayer metal; the dashed line in both graphs represents linear fit to extract the excess current. Conductance spectra for three junctions with (\textbf{b}) - Ti and (\textbf{d}) - Pt interlayer metals; the arrows indicate the deeps corresponding to $2\Delta_\mathrm{S'}$.}
\label{fig:FigS4}
\end{figure}

\ref{fig:FigS4} compares the conductance spectra, obtained by differentiating the current voltage characteristics (IVCs), of various junctions fabricated with Ti interlayer, with those obtained with a Pt interlayer. 
The curves of \ref{fig:FigS4}b (Ti interlayer) clearly show a dip at $V \approx 230 \, \mu$eV  that we  have identified as   $2\Delta_\mathrm{S'}$, where  $\Delta_\mathrm{S'}$ is the induced gap into the TI \cite{kjaergaard2016, galletti2017high}.  In the spectra of \ref{fig:FigS4}b no other dips can be clearly identified.  \ref{fig:FigS4}d shows instead various spectra relative to the IVCs of junctions fabricated with Pt interlayer. Also in this case we can identify a well developed dip at $V \approx 230 \, \mu$eV  (much more pronounced than the Ti case due to the higher transparency $\tau$ of the barrier I) that we correlate with  $2\Delta_\mathrm{S'}$, where again  $\Delta_\mathrm{S'}$  is the induced gap (this correlation is unambiguous because of the peculiar temperature dependence of  $2\Delta_\mathrm{S'}$ \cite{kjaergaard2016, galletti2017high}).

However the conductance curves of the junctions with Pt have many more dips at different voltages, associated to multiple Andreev reflections and that are made visible  because of the higher transparencies interfaces I, as we find in our experiment\cite{galletti2017high}.  To summarize these measurements we observe 1) the interface Bi$_2$Te$_3$/Pt/Al and Bi$_2$Te$_3$/Ti/Al have the same transparency since the induced gap  $\Delta_\mathrm{S'} \approx 115 \, \mu$eV  is {\itshape exactly the same for both kind of junctions}  and 2)  the  difference in the $J_\mathrm{C}$ values, $I_\mathrm{C}R_\mathrm{N}$  and transparency $\tau$ has to be related to the higher  transparent interfaces I, between the part of the flake in the nano-channel and that under the electrodes,  for the Pt junctions.     The magnetic pattern of the Josephson current for both Pt and Ti  junctions are Fraunhofer-like at the first cool down. However at the second cool down the junctions with Ti interlayer are not affected by the thermal cycle. The IVCs and the magnetic Fraunhofer pattern do not have any variation. We have cooled down the junctions from room to  20 mK   temperature 3 times and the transport properties are completely preserved.  This fact is in line with the results obtained by other groups working with 3DTI and using Ti as sticking layer.

\vspace{0.5cm}
{\itshape What makes the junctions with Pt interlayers so different from the Ti?}
\vspace{0.5cm}

We tend to  exclude  an origin related to a different chemistry  between Pt and Ti while bonding to the Bi$_2$Te$_3$. This is because the values of the induces gap $\Delta_{S'}$ are identical, which is a consequence of very similar interface transparency.

Instead we will show that the completely different behavior between Ti and Pt junctions is related to different growth habits of the two metals on Bi$_2$Te$_3$.

\begin{figure}[!ht]
\includegraphics[width=15cm]{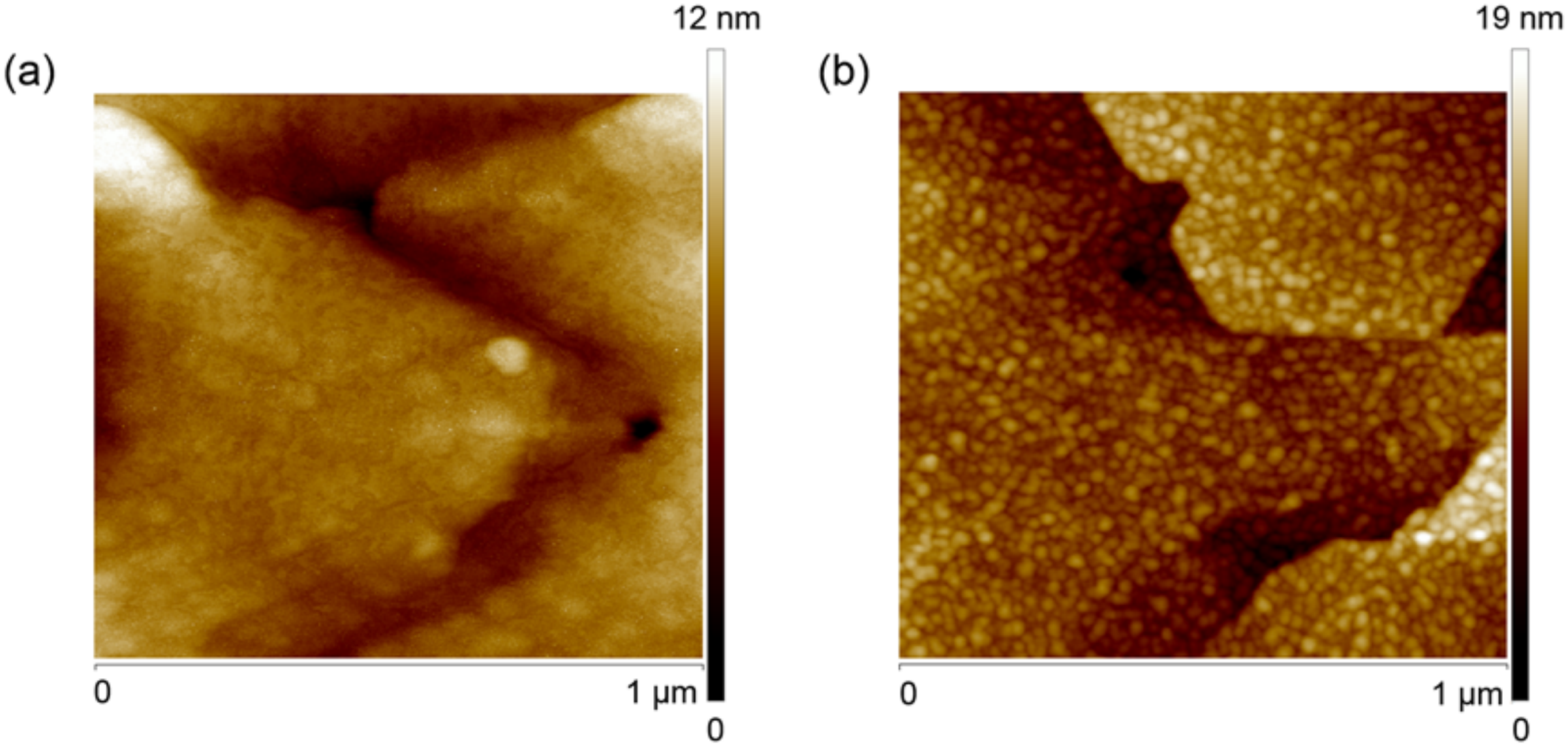}
\caption{\textbf{Growth habits of Pt and Ti on a  Bi$_2$Te$_3$ flake.} AFM images of a Bi$_2$Te$_3$ flake covered (\textbf{a}) with 3 nm Pt and (\textbf{b}) with 3 nm Ti. }
\label{fig:FigS5}
\end{figure}

\ref{fig:FigS5}a shows an AFM picture of a 3 nm Pt grown on a Bi$_2$Te$_3$ flake. The image shows a uniform layer that nicely wets the Bi$_2$Te$_3$ flake smoothening edges and corners. The growth habits of  a 3 nm Ti film are instead quite different (see \ref{fig:FigS5}b). 
The Ti  does not wet forming tiny grains (on average 20 nm in diameter) on the surface of Bi$_2$Te$_3$.  As a result the morphology of the Al film, which grows on top of the Ti is much more granular (data not shown), compared to the growth on Pt. 

\begin{figure}[b]
\includegraphics[width=16cm]{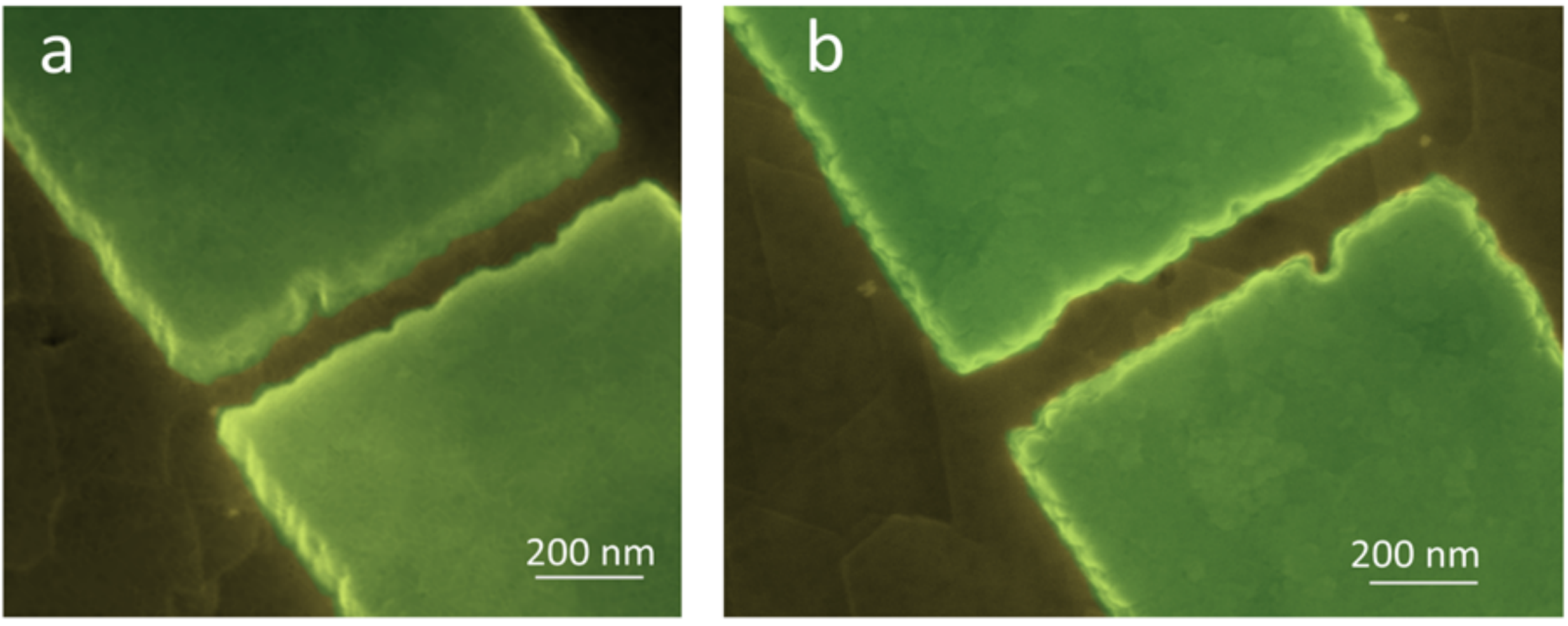}
\caption{\textbf{Colored SEM images for a junction  with Al electrodes and  Ti interlayer.} (\textbf{a}) Tilted view; (\textbf{b})  side view. The electrodes are represented in light green color.}
\label{fig:FigS6}
\end{figure}

The Al electrodes have the role of clamping the flake to the substrate so as to experience a tensile and compressive strain. For this to happen the Al and the flake underneath need to behave like almost the same material. The wetting properties of the Pt on the Bi$_2$Te$_3$  appear to promote an excellent  Al  adhesion and therefore an effective clamping of the flake to the substrate. The granular growth of the Ti interlayer on Bi$_2$Te$_3$  instead does not seem to be  as effective in this respect.  While the interface properties of the Al  and Bi$_2$Te$_3$  through the Ti are very similar to the Pt junctions, the granularity of the Ti sticking layer  makes the Al and the Bi$_2$Te$_3$ flake  not as strongly connected to each other to behave as a homogenous materials. In this way  no  plastic deformations  are induced. \ref{fig:FigS6}b shows a typical  top view SEM picture  for  one of the measured junction after several cool down; \ref{fig:FigS6}a is a SEM image of the same device taken with a tilted angle. None of the images show signature  of buckling wave  that instead we have detected in junctions with Pt interlayer. 

It is worth mentioning what  could be the origin of the much higher critical current densities we observe in junctions with Pt interlayer compared with those with Ti at the first cool down.  We believe  it is a  consequence of the occurrence of a tensile strain  experienced by the flake in the nanogap. 

The origin of the barrier I between the flake under the Al/Pt(Ti) electrodes and the one in the nanogap is not clearly established. One can  possibly expect some doping from the metal, which shifts the Fermi energy of the flake under the electrode, compared to the one in the nanogap. The occurrence of  a tensile strain at the nanogap while cooling the nanodevice, shift the Dirac point toward the valence bands making it more doped. Since the effect of strain is much more dominant in  junctions with Pt, one can expect a better Fermi level matching  between the two parts of the flake, under the electrodes and at the nanogap, resulting in a higher transparency barrier and higher $J_\mathrm{C}$ values.

\section{Two dimensional Magneto-fingerprints in \NoCaseChange{Bi$_2$Te$_3$}  channels}

In the paper by  Kandala et al. \cite{kandala2013surface}, the authors study the magneto transport of Bi$_2$Se$_3$ channels, where the films have a typical growth with pyramidal domains. The authors reveal signatures of Aharonov-Bohm (AB) orbits, manifesting as periodic  magneto conductance fluctuations. The length scale of the orbits corresponds to the typical perimeter of triangular terraces found on the surface of these thin film devices. They conclude that the periodic magneto-fingerprint arises from coherent scattering of electron waves from the step-edges. To demonstrate that this scenario is also valid for the Bi$_2$Te$_3$ flakes, used in our work, we have preformed magneto transport measurements  down to 20 mK temperature on devices with a Hall configuration. 

\begin{figure}[!ht]
\includegraphics[width=12.2cm]{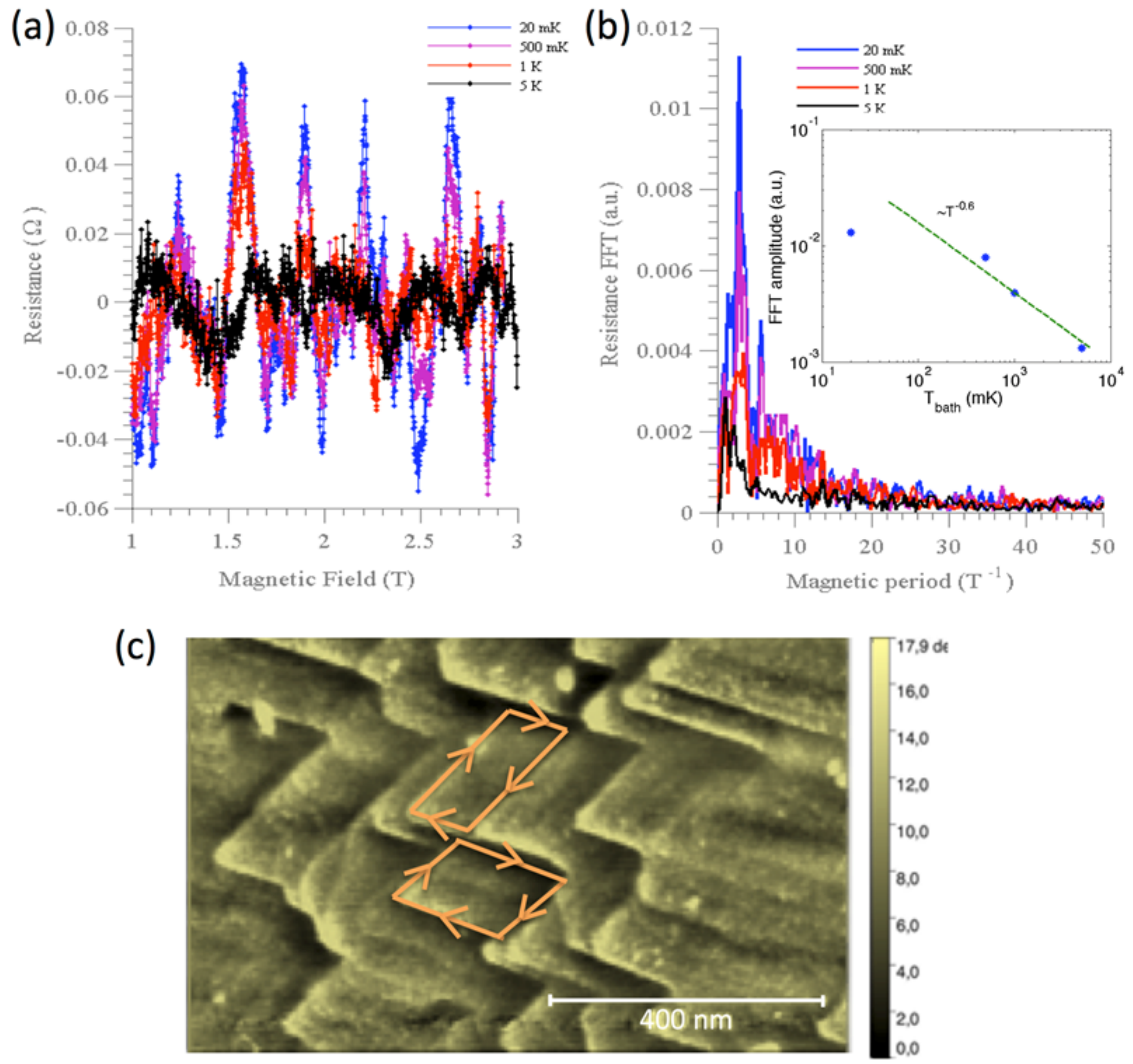}
\caption{\textbf{Two dimensional magneto-fingerprints in Bi$_2$Se$_3$ channels.} \\ (\textbf{a}) Longitudinal resistance  as a function of magnetic field  for a channel length of 0.5 $\mu$m  at different temperatures. (\textbf{b}) Fourier transform of the curves of panel (a). The inset shows the Fourier transform amplitude at  2.9 T$^{-1}$  as a function of temperature. The data have a typical power law dependence (with exponent -0.6). (\textbf{c}) Typical surface morphology of the flakes used in the experiment. The orange lines delimitate two possible Arhonov-Bhom orbits with approximately the same area.  }
\label{fig:FigS7}
\end{figure}

\ref{fig:FigS7}a  shows the longitudinal resistance between the two voltage  probes with  a distance of 0.5 $\mu$m (the  classical quadratic magneto-resistance background has been subtracted for clarity) in the range of magnetic field 1-3 T  for various temperatures. From the Fourier transform of the curves, \ref{fig:FigS7}b, we clearly observe a peak around  2.9 T$^{-1}$  which corresponds to an area of  0.012 $\mu$m$^2$ in line with the morphology of our flakes  and in complete agreement with the values reported in Ref. \citenum{kandala2013surface} for a device of similar dimensions  as ours.  Moreover we clearly see that the amplitude of the magnetoresistance oscillations decreases with temperature  (inset of \ref{fig:FigS7}b) in a similar fashion as in Ref. \citenum{kandala2013surface}.  This strongly suggests that also in our case the periodic magneto-resistance oscillations arise from coherent scattering of electron wave from the corners  and/or step-edges where two pyramidal domains merge. 

\begin{figure}[!ht]
\includegraphics[width=11.5cm]{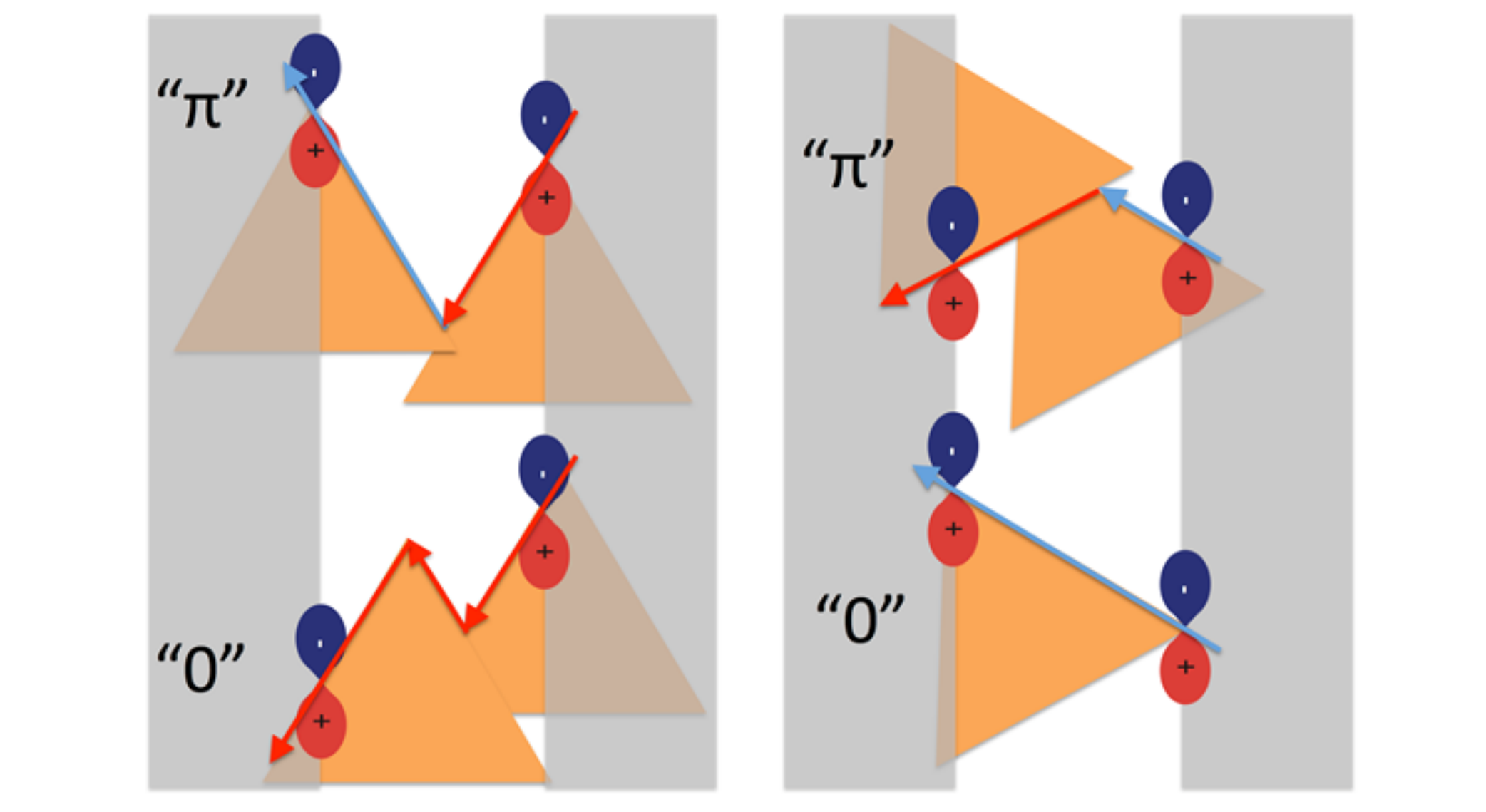}
\caption{\textbf{Schematic pictures of a Josephson junction with different orientation of the pyramidal domains in the nanogap with respect to the Al electrodes.} In the left panel, the main direction of the pyramidal domains is aligned parallel to the Al electrodes, while it is perpendicular to them in the right panel. In the figure, 0 and $\pi$ trajectories can be found in both cases.}
\label{fig:FigS8}
\end{figure}

It is also worth discussing if one can find any apparent correlation, between the orientation of the nano-pyramidal domains and that of the electrodes, with the occurrence of unconventional magnetic field pattern related to 0 and $\pi$ trajectories inside the Bi$_2$Te$_3$ nanochannel. We have investigated  by SEM all the junctions having a 0-$\pi$ transition and we do not find any clear correlation between the orientation of the nano-pyramidal domains and that of the electrodes.   
However this is not surprising if one considers that the basic mechanism to get an inverted magnetic field pattern, with a minimum of the critical current at  zero external field, is the occurrence of 0 and $\pi$ facets within the same nanogap. In our flakes the occurrence of AB oscillations supports preferential trajectories of electrons and holes along the edges of  the triangles.  
\ref{fig:FigS8} 
shows a sketch  corresponding to two extremes cases where the main alignment direction of the pyramidal domains is parallel (\ref{fig:FigS8}a)  or perpendicular (\ref{fig:FigS8}b)  to the electrodes.   We see that in both cases one can find 0 and $\pi$ trajectories, which would lead to an inverted Fraunhofer pattern. 

\section{Periodicity  of a magnetic pattern  with  regular and random distributed  0-$\pi$ facets.}

\begin{figure}[p]
\includegraphics[width=8cm]{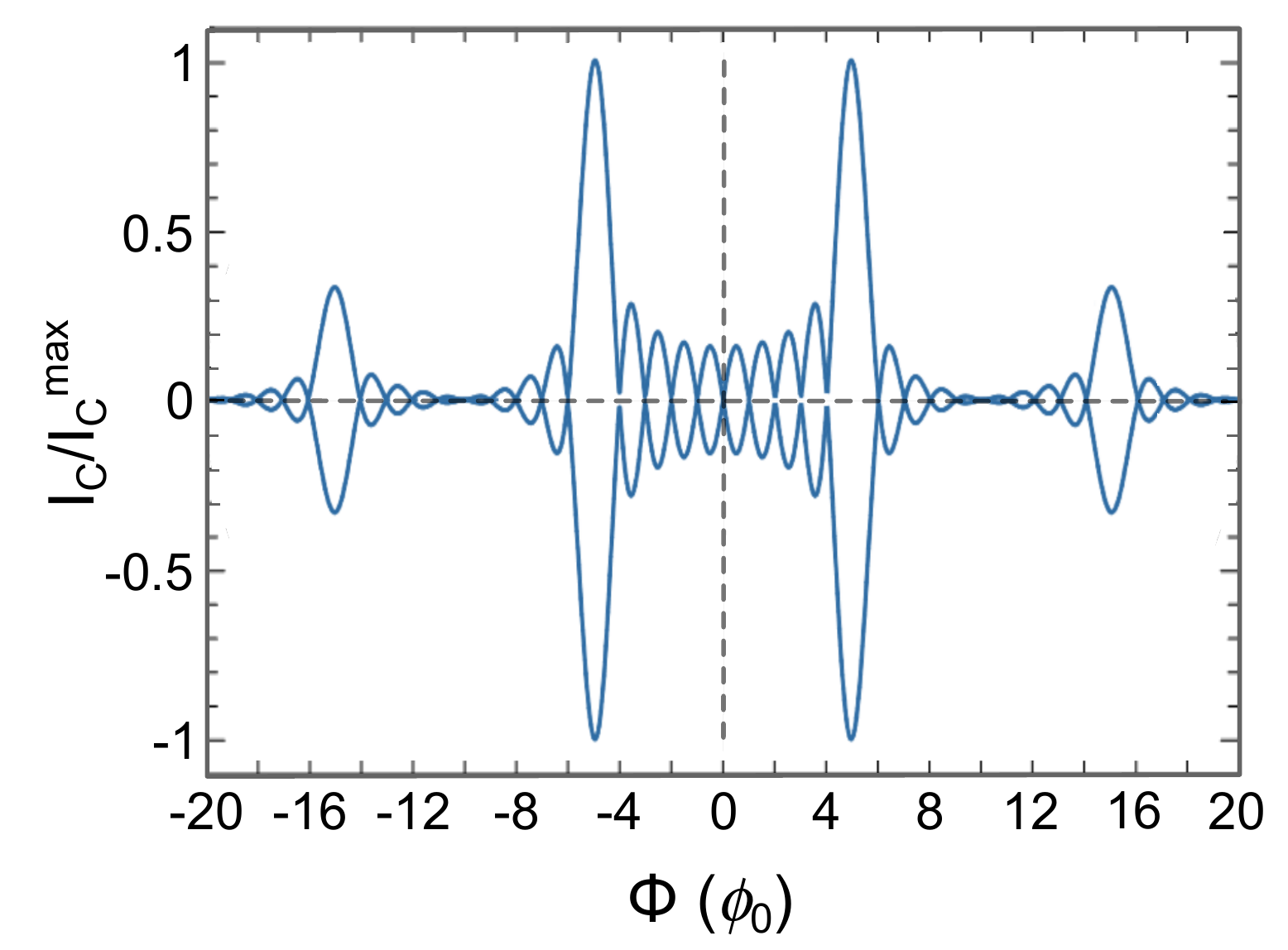}
\caption{\textbf{Simulated magnetic flux dependence of the Josephson current for an array of  10 alternating 0 and} $\bm{\pi}$ \textbf{facets.}}
\label{fig:FigS10}
\end{figure}

The magnetic patter of a regular 0-$\pi$ faceted Josephson junction is strongly modified compared to  a Fraunhofer-like. \ref{fig:FigS10} 
shows the $I_\mathrm{C}(B)$ for a junction formed by  a total of 10  equal in size 0 and $\pi$ facets  carrying the same current \cite{smilde2002dwave}. The absolute maximum of the pattern is found at 5$\phi_0$. This is quite easy to understand since one requires a $\phi_0$  to reverse the  phase sign of  every $\pi$ facet.   However one can still find the periodicity connected to the total junction's width in the position of the minima which happen at multiple of $\phi_0$  (see \ref{fig:FigS10}). For a random distribution of facets the absolute maximum will always occur at finite field (which will depend on the microscopic distribution of  0-$\pi$ facets), while at zero external field one can find a local maximum (not the absolute maximum) or a local minimum again dependent on the details of  the facets microscopic distribution.

\begin{figure}[p]
\includegraphics[width=13cm]{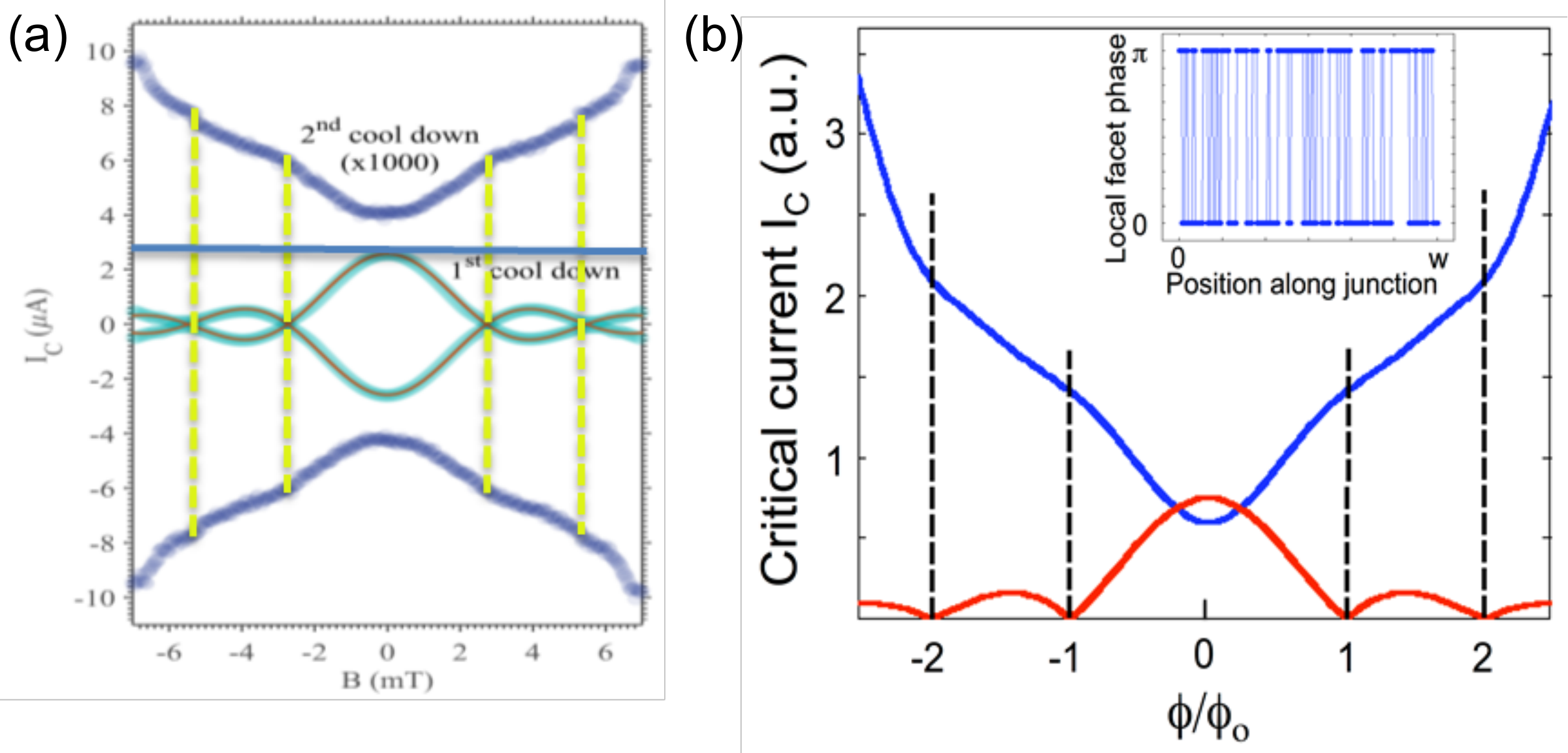}
\caption{\textbf{Magnetic field dependence of a typical Bi$_2$Te$_3$ junction compared with that of a simulated array of  random distributed 0 and} $\bm{\pi}$ \textbf{facets.} \\ (\textbf{a}) Critical current $I_\mathrm{C}$ dependence on the externally applied magnetic field $B$ for  one of 
our junctions at 20 mK, before (cyan) and after (blue) the thermal cycle. At the first cool down the $I_\mathrm{C}(B)$ dependence shows a conventional Fraunhofer dependence. After the thermal cycle (blue points), the critical current is dramatically reduced (the data points are multiplied by a factor 1000 for clarity), and a dip at $B$=0 appears. (\textbf{b}) Numerically calculated $I_\mathrm{C}(B)$ (blue curve) for a random distribution of 0 and $\pi$  facets (see inset) compared with a conventional Fraunhofer type dependence calculated considering a uniform current distribution.
}
\label{fig:FigS11}
\end{figure}

The periodicity connected to the size of the junction can be still identified as local minima, local maxima or change in slope in the $I_\mathrm{C}(B)$.  As an example in \ref{fig:FigS11}(b) we show the computed magnetic field dependence  $I_\mathrm{C}(B)$ for  a uniform  current distribution (red curve) compared with a random distribution of   0 and $\pi$ facets (blue curve). 

For the random case we can clearly identify change of slopes at $\phi_0$ and 2$\phi_0$  (indicated by the dashed lines) which corresponds to the periodicity determined by the width of the junction. Indeed such a scenario qualitatively reproduces what we observe in our experiment (see \ref{fig:FigS11}(a)) and that is discussed in the manuscript.

\section{Fit of the $\bm{I_\mathrm{C}(T)}$ dependence with a SINIS model}

The full theory of the Josephson effect in SINIS structures with chiral p-wave ($p_x$ + i$p_y$) superconducting leads is still lacking. In Ref. \citenum{sawa2007quasiclassical}, Sawa et al have considered a SINIS system with $p_x$ + i$p_y$ superconducting leads and diffusive normal part, numerically solved Usadel equations and determined the $I_\mathrm{C} (T)$ dependence. This dependence turns out to be qualitatively very similar to that of an usual SINIS junction with s-wave leads. Moreover, in the limit of highly transparent insulating barriers $I_\mathrm{C} (T)$ dependence for $s$-wave and $p$-wave leads are found to be essentially identical. In Ref. \citenum{tkachov2013helical} the authors have considered a SINIS structure with chiral $p$-wave leads and short ballistic normal metal. These authors have also found that for perfectly transmitting barriers both $p_x$ + i$p_y$ and $s$-wave pairing symmetries lead to the same $I_\mathrm{C} (T)$ dependence. In view of these previous works, we have modeled our highly transparent junctions measured during the first cool down as S'INIS' structures (where S' is the proximity induced superconductor) with conventional s-wave superconducting leads, for which a simple analytical formula for the Josephson current exists in the clean ballistic limit \cite{galaktionov2002quantum}. The same approach has been adopted in an earlier work \cite{veldhorst2012josephson}.

For our S'INIS' junctions, we consider the limit of wide junction, $k_\mathrm{F}w \gg 1$, and use the following expression for the critical current 
\begin{eqnarray}
I_J(\phi) = \frac{4ek_\mathrm{B}T}{\hbar} \frac{k_\mathrm{F}w}{\pi} \sin\phi \sum\limits_{\omega_n>0} \int_0^1 d\mu \frac{t_1t_2}{\sqrt{Q\left(\phi, \sqrt{1-\mu^2}\right)}} \; .
\label{eq:eq5}
\end{eqnarray}
Here $\mu = \sin\theta$, where $\theta$ is the angle between the velocity of an electron flying out of a lead and the shortest line connecting the two leads,
\begin{eqnarray}
t_1 = \frac{D_1}{2 - D_1} \;, \;\;\;\;\;\; t_2 = \frac{D_2}{2 - D_2}
\label{eq:eq6}
\end{eqnarray}
are the effective Andreev transparencies of the barriers\cite{veldhorst2012josephson}, $D_1$ and $D_2$ are the usual barrier transparencies in the normal state. In Eq. (\ref{eq:eq5}), the function $Q$ is defined as 
\begin{equation}
\begin{multlined}
Q\left(\phi, \sqrt{1-\mu^2}\right) = \left[t_1t_2\cos\phi + \left(1+(1+t_1t_2)\frac{\hbar^2\omega_n^2}{\Delta^2}\right)\cosh\left[2\omega_nt_0(\mu)\right] \right. + \\ \left. +  (t_1+t_2)\frac{\hbar^2\Omega_n\omega_n}{\Delta^2}\sinh\left[2\omega_nt_0(\mu)\right] \right]^2 - (1-t_1^2)(1-t_2^2)\frac{\hbar^4\Omega_n^4}{\Delta^4} \; ,
\label{eq:eq7}
\end{multlined}
\end{equation}
\begin{eqnarray}
\omega_n = \frac{\pi k_\mathrm{B}T(2n+1)}{\hbar} \; ,
\label{eq:eq8}
\end{eqnarray}
\begin{eqnarray}
\hbar\Omega_n = \sqrt{\hbar^2\omega_n^2 + \Delta^2} \; ,
\label{eq:eq9}
\end{eqnarray}
$\omega_n$  are the Matsubara frequencies, and $t_0(\mu)$ is the angular dependent average flight time of an electron between the leads for a given Fermi speed $v_\mathrm{F}$, mean free path $l_\mathrm{e}$, the separation between the leads $L$, and
\begin{eqnarray}
t_0(\mu) = \frac{l_e}{(1-\mu^2)v_\mathrm{F}\left(\sqrt{1+\frac{l_\mathrm{e}^2}{L^2(1-\mu^2)}}-1\right)} \; .
\label{eq:eq10}
\end{eqnarray}
Supplementary Equation (\ref{eq:eq10}) may be viewed as an interpolation between the clean limit ($l_\mathrm{e}  \gg L$), in which case 
\begin{eqnarray}
t_0(\mu) = \frac{L}{v_\mathrm{F}\sqrt{1-\mu^2}} \; ,
\label{eq:eq11}
\end{eqnarray}
and the diffusive limit ($l_\mathrm{e} \ll L$) , where $t_0$ does not any more depend on the angle and reads
\begin{eqnarray}
t_0 = \frac{2L^2}{v_\mathrm{F}l_\mathrm{e}} \; .
\label{eq:eq12}
\end{eqnarray}
For the normal state resistance in this model we find
\begin{eqnarray}
\frac{1}{R} = \frac{e^2}{\pi\hbar} \frac{k_\mathrm{F}w}{\pi} \frac{1}{\frac{1}{D_1}+\frac{1}{D_2}-1+\frac{2L}{\pi l_\mathrm{e}}} \; .
\label{eq:eq13}
\end{eqnarray}

The model outlined above differs from the original one\cite{galaktionov2002quantum} in two ways. Since we are dealing with a two dimensional normal layer (while in Ref. \citenum{galaktionov2002quantum} a three dimensional normal metal was considered), we have replaced the integral over the two transverse components of the wave vector, $k_y$ and $k_z$, by the corresponding one-dimensional integral over $k_y$, i.e. we have replaced $dk_y dk_z/(2\pi)^2$ by $dk_y/2\pi h$, where $h$ is the flake thickness. Second, we have introduced finite mean free path into the expression for the Josephson current in a phenomenological way. 

Namely, we have replaced the ballistic expression for the flight time, Supplementary Equation (\ref{eq:eq11}), by a more complicated one containing the mean free path, Supplementary Equation (\ref{eq:eq10}). We have verified that this replacement correctly reproduces the known expression for the Josephson current of a short diffusive SINIS junction.

\begin{figure}[!ht]
\includegraphics[width=14cm]{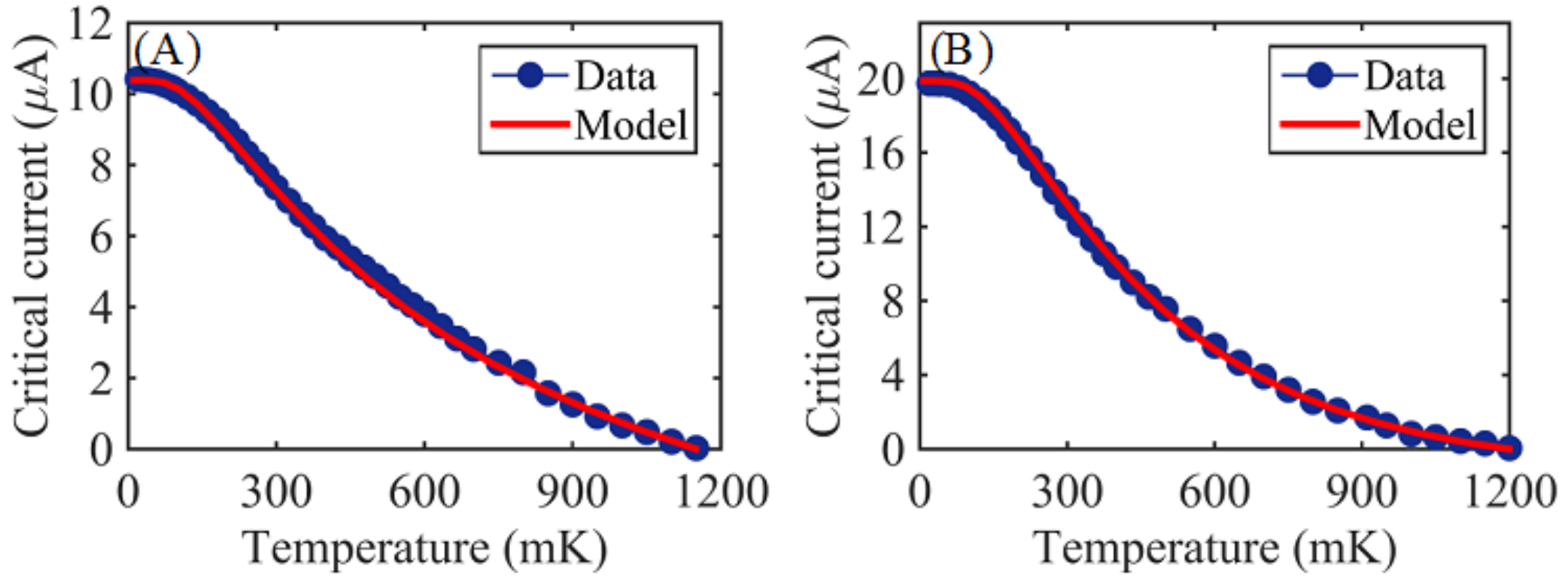}
\caption{\textbf{Critical current as a function of the temperature at the first cool down.} The measured $I_\mathrm{C}(T)$ dependencies are for samples MS1 (panel \textbf{A}) and FS4 (panel \textbf{B}). The red line is the best fit, considering a quasi-ballistic model (Supplementary Equations (\ref{eq:eq5}) and (\ref{eq:eq7})). The fit parameters are listed in \ref{tab3}.}
\label{fig:FigS9}
\end{figure}

We have performed a fit on a selection of the junction listed in Table 1 (see main text), which covered a wide range in the magnitude of the critical current values (see \ref{fig:FigS9}). We used a value for the S' superconducting gap of $\Delta_\mathrm{S'}  \approx 125 \,\mu$eV, which is close to the value of the Al superconducting gap.
The Fermi velocity of the electrons has be chosen to be $v_\mathrm{F} = 3.5 \cdot 10^5$ m s$^{-1}$, in agreement with literature, and the fitting parameters are: the Fermi energy, $E_\mathrm{F}$, the mean free path, $l_\mathrm{e}$, and the transparencies of the barriers, which we assumed to be identical, $D_1 = D_2 = D$. The theoretical value of the normal state resistance, $R_\mathrm{th}$, has been determined from Supplementary Equation (\ref{eq:eq13}) and the Thouless energy was estimated as $E_\mathrm{Th} =hv_\mathrm{F} l_\mathrm{e}/L^2$.

The extracted values of the mean free path ($l_\mathrm{e}$) are typically comparable to the junction length, thus confirming a picture of an intermediate, between ballistic and diffusive, transport regime. The scattering centers are probably the edges of the pyramidal domains of the film, as we have discussed in the main text.

It is known\cite{veldhorst2012josephson} that the bulk of the flake should behave as a resistive shunt. From the expression
\begin{eqnarray}
\frac{1}{R_\mathrm{sh}} = \frac{1}{R_\mathrm{exp}} - \frac{1}{R_\mathrm{th}}      
\label{eq:eq14}
\end{eqnarray}
we have extracted the effective shunt resistances of the devices. In Supplementary Equation (\ref{eq:eq14})  $R_\mathrm{th}$ is the prediction of Supplementary Equation (\ref{eq:eq13})  and $R_\mathrm{exp}$ is the resistance measured in the experiment. The sheet resistances of the films associated with the bulk transport, $R_\square = R_\mathrm{sh} w/L$, vary in the range between 150 and 550 $\Omega/\square$. The corresponding resistivities, defined as $\rho = R_\square h$, are shown in \ref{tab3} and change between 1000 and 5000 $\mu\Omega$cm, which is about one order of magnitude larger than the bulk resistivity reported in Ref. \citenum{veldhorst2012josephson}, which probably indicates a lower level of doping in our material. We believe that the scattering of the extracted values of the bulk resistivity between different devices is mostly caused by the uncertainty in the geometry of the samples, but it may also be explained by slightly different levels of doping.

\begin{table}[hbpt!]
\resizebox{10cm}{!}{
\begin{tabular}{| c || c | c | c || c | c | c | c | c | }
\hline
 \textbf {Device}  & \textbf {$l_\mathrm{e}$}  &$D$ &$E_\mathrm{F}$&$E_\mathrm{Th}$&$R_\mathrm{N}^{\mathrm{th}}$&$R_\mathrm{sh}$&$R_\square$& $\rho$  \\
 & (nm) & &(meV)&(${\mu}$eV)&($\Omega$)&($\Omega$)&($\Omega$)&($\mu\Omega$cm)  \\
\hline
 
\textbf {JM1} & 140 & .99 &106&363&566    &101 &152 &1200 \\
\hline
\textbf {JM10} & 130 & .98 &140&341&134    &47 &236 &1900 \\
\hline
\textbf {SM4} & 100 & .98 &172&954&45    &27 &543 &4300 \\
\hline
\textbf {SM1} & 95 & .98 &213&920&37    &25 &502 &4000 \\
\hline
\textbf {SF4} & 35 & .94 &210&392&22    &3.5 &209 &1670 \\
\hline
\end{tabular}}
\caption{\textbf{Parameters extracted from the fit of the} \bm{$I_\mathrm{C} (T)$}. The mean free path $l_\mathrm{e}$, the transparency of the barriers $D$ (assuming the two barriers are identical) and the Fermi energy $E_\mathrm{F}$ are the free parameters of the fit. From these parameters we derived the Thouless energy $E_\mathrm{Th}$, the expected resistance above the gap $R_\mathrm{N}^{\mathrm{th}}$, the calculated shunt resistance of the flake $R_\mathrm{sh}$, and the corresponding sheet resistance $R_\square$ and resistivity $\rho$. The induced gap at zero temperature was taken to be $\Delta(0)=125 \, \mu$eV and at higher temperatures the standard BCS gap temperature dependence has been used.  \label{tab3}} 
\end{table}

\newpage


\end{document}